\documentclass[usegraphicx,useAMS,usenatbib]{mn2e}
\usepackage{xspace}
\usepackage[T1]{fontenc}
\usepackage{verbatim}
\usepackage[normalem]{ulem}
\usepackage{amsfonts,amsmath,amssymb}
\usepackage{times}
\usepackage[italic]{mathastext}
\usepackage{enumitem}
\usepackage[colorlinks,linkcolor=blue,citecolor=blue,urlcolor=blue,breaklinks]{hyperref}
\usepackage{color}
\usepackage{upgreek}
\usepackage{flushend}
\usepackage{lineno}
\usepackage{siunitx}
\usepackage{caption}

\usepackage{eso-pic}% http://ctan.org/pkg/eso-pic

\AddToShipoutPictureBG*{%
  \AtPageUpperLeft{%
    \hspace{10pt}%
    \raisebox{-5\baselineskip}{%
      \makebox[0pt][l]{\textnormal{DES-2018-0372}}
}}}%

\AddToShipoutPictureBG*{%
  \AtPageUpperLeft{%
    \hspace{10pt}%
    \raisebox{-6\baselineskip}{%
      \makebox[0pt][l]{\textnormal{FERMILAB-PUB-18-626-AE}}
}}}%

\def\reff@jnl#1{{\rm#1\/}}
\def\aj{\reff@jnl{AJ}}         % Astronomical Journal
\def\araa{\reff@jnl{ARA\&A}}      % Annual Review of Astron and Astrophys
\def\apj{\reff@jnl{ApJ}}        % Astrophysical Journal
\def\apjl{\reff@jnl{ApJ}}        % Astrophysical Journal, Letters
\def\apjs{\reff@jnl{ApJS}}       % Astrophysical Journal, Supplement
\def\aap{\reff@jnl{A\&A}}        % Astronomy and Astrophysics
\def\aapr{\reff@jnl{A\&A~Rev.}}     % Astronomy and Astrophysics Reviews
\def\aaps{\reff@jnl{A\&AS}}       % Astronomy and Astrophysics, Supplement
\def\mnras{\reff@jnl{MNRAS}}      % Monthly Notices of the RAS
\def\physrep{\reff@jnl{Physics Reports}}% Physics Reports
\def\prd{\reff@jnl{Phys.Rev.D}}     % Physical Review D
\def\prl{\reff@jnl{Phys.Rev.Lett}}   % Physical Review Letters
\def\pasp{\reff@jnl{PASP}}       % Publications of the ASP
\def\pasj{\reff@jnl{PASJ}}       % Publications of the ASJ
\def\nat{\reff@jnl{Nature}}       % Nature
\def\jcap{\reff@jnl{JCAP}}   %Journal of Cosmology and Astroparticle Physics
\def\memsai{\reff@jnl{MemSAI}} %Memorie della Societa Astronomica Italiana Supplement 
\def\na{\reff@jnl{New Astronomy}}       % New Astronomy
\def\procspie{\reff@jnl{SPIE}}       %SPIE 
\def\pasa{\reff@jnl{PASA}}       % Publication of the Astronomical Society of Austrailia

\def\Fref#1{Fig.~\ref{#1}\xspace}

\def\Cref#1{Chapter~\ref{#1}\xspace}

\title[Splashback in SZ-selected Clusters with DES galaxies]{Measurement of the Splashback Feature around SZ-selected Galaxy Clusters with DES, SPT and ACT}

\author[Shin et al.]
{\parbox{\textwidth}{
\Large
T.~Shin,$^{1}$\thanks{E-Mail: taeshin@sas.upenn.edu}
S.~Adhikari,$^{2,3}$
E.~J.~Baxter,$^{1}$
C.~Chang,$^{4,5}$
B.~Jain,$^{1}$
N.~Battaglia,$^{6}$
L.~Bleem,$^{4,7}$
S.~Bocquet,$^{4,7,8}$
J.~DeRose,$^{2,9}$
D.~Gruen,$^{2,3}$
% ...
  \vspace{0.2cm}\\
  \parbox{\textwidth}{
(affiliations are listed at the end of the paper)\\}}
}
\begin{document}
\date{\today}
\pagerange{\pageref{firstpage}--\pageref{lastpage}}
\pubyear{2018}
\maketitle
\label{firstpage}

\begin{abstract}
We present a detection of the splashback feature around galaxy clusters selected using the Sunyaev-Zel'dovich (SZ) signal.
Recent measurements of the splashback feature around optically selected galaxy clusters have found that the splashback radius, $r_{\rm sp}$, is smaller than predicted by N-body simulations.
A possible explanation for this discrepancy is that $r_{\rm sp}$ inferred from the observed radial distribution of galaxies is affected by selection effects related to the optical cluster-finding algorithms.
We test this possibility by measuring the splashback feature in clusters selected via the SZ effect in data from the South Pole Telescope SZ survey and the Atacama Cosmology Telescope Polarimeter survey. The measurement is accomplished by correlating these cluster samples with galaxies detected in the Dark Energy Survey Year 3 data. 
The SZ observable used to select clusters in this analysis is expected to have a tighter correlation with halo mass and to be more immune to projection effects and aperture-induced biases, potentially ameliorating causes of systematic error for optically selected clusters.  
We find that the measured $r_{\rm sp}$ for SZ-selected clusters is  consistent with the expectations from simulations, although the small number of SZ-selected clusters makes a precise comparison difficult. In agreement with previous work, when using optically selected redMaPPer clusters with similar mass and redshift distributions, $r_{\rm sp}$ is $\sim$$2\sigma$ smaller than in the simulations.  These results motivate detailed investigations of selection biases in optically selected cluster catalogs and exploration of the splashback feature around larger samples of SZ-selected clusters.
Additionally, we investigate trends in the galaxy profile and splashback feature as a function of galaxy color, finding that blue galaxies have profiles close to a power law with no discernible splashback feature, which is consistent with them being on their first infall into the cluster.
\end{abstract}


\section{Introduction}
Halos are non-linear, gravitationally bound structures where dark matter particles are in orbits governed by the halo gravitational potential, detached from the overall expansion of the universe.
The physical boundary of a dark matter halo is the surface corresponding to the largest apocenters of the material that has been accreted into the halo most recently.
This forms a phase space boundary between the outer regions where objects are on first infall and the region within a halo where dark matter is  ``virialized'' or multistreaming, i.e. orbiting shells of dark matter are crossing each other leading to multiple streams at a given point.
This boundary is clearly visible in the outskirts of simulated dark matter halos as a sharp decline in the slope of the density profile, and the location at which the slope reaches a minimum is called the \textit{splashback radius}, $r_{\rm sp}$ \citep{Diemer2014,Adhikari2014, mansfield2017,Diemer2017,Okumura2017,Okumura2018}. While a density caustic feature at the boundary of a halo at first turnaround after infall was suggested by theoretical work based on the smooth spherical collapse models \citep[e.g.][]{Gunn1972,Fillmore1984,Bertschinger1985,Adhikari2014,2016MNRAS.459.3711S}, \citet{Diemer2014} presented evidence that this feature appears in the profiles of realistically simulated dark matter halos, even after averaging over halos of different masses, accretion histories, and redshifts.  

The profiles of actual dark matter halos in the universe can be probed in several ways, for example, by studying
the distribution of galaxies in halos, which is determined by the gravitational potential of the overall matter distribution, or by stacking the weak gravitational lensing of background galaxies around halos to get the matter distribution directly. \citet{More2016} used the galaxy surface density profile around redMaPPer (RM) \citep{Rykoff2014} galaxy clusters identified in data from the Sloan Digital Sky Survey \citep[SDSS,][]{Aihara2011} to present the first evidence for a splashback feature. Subsequently, evidence for the feature was also found by \citet{Baxter2017} using the galaxy surface density profiles around two samples of SDSS-identified clusters, and by \citet{Chang2017} using the galaxy density and weak lensing profiles around RM clusters identified in the first year of Dark Energy Survey (DES) data.
Recently, \citet{contigiani2018} has measured weak lensing profiles around 27 massive clusters obtained with the Cluster Canadian Comparison Project (CCCP) and reported a measurement of the splashback radius.  However, \citet{contigiani2018} do not report a statistically significant detection of splashback-like steepening in the cluster density profile (see also \citealt{diemerlens2017} for lensing-based constraints on $r_{\rm sp}$ with clusters from the Cluster Lensing And Supernova survey with Hubble (CLASH)).
In all of these cases, the evidence for the splashback feature came from identifying the presence of a sharp steepening in the halo (galaxy/dark matter) surface density profiles. Interestingly, for clusters identified via the RM algorithm, and for measurements using the galaxy surface density profile around clusters, the location of splashback is about 20\% ($\sim$$3\sigma$) smaller than predictions from N-body simulations \citep{More2016,Baxter2017,Chang2017}.

\citet{Busch2017} explored whether cluster-finding algorithms like RM can imprint artificial splashback-like features into cluster density profiles via selection effects. In essence, the problem arises due to selecting halos based on cluster richness, $\lambda$, which for RM is measured within an aperture, $R_{\lambda}=1.0(\lambda/100)^{0.2}h^{-1}{\rm Mpc}$. Clusters with galaxies just inside $R_{\lambda}$ are more likely to be included in the richness selected sample than clusters with galaxies just outside $R_{\lambda}$. So a feature associated with the selection aperture due to random fluctuations of the galaxy distribution relative to the dark matter may be imprinted on the profile. 
\citet{Zu2017} and \citet{Busch2017} have also pointed out that projection effects in the RM catalog can impact the amplitude of 2D cluster-galaxy correlations at large scales. Because RM identifies clusters with imaging data, some fraction of galaxies identified as cluster members may actually be chance projections of background galaxies along lines of sight near to the cluster. \citet{Zu2017} showed that projections are more likely to occur in dense regions, causing the cluster concentration inferred from member galaxy positions to correlate with large scale overdensities.
\citet{Busch2017}, however, concluded based on mock-RM simulations that while projections could bias the inference of cluster member concentration, projections did not alter the location of the splashback feature averaged over all clusters.

\citet{Baxter2017} investigated the impact of potential RM systematic effects on measurements of the splashback feature with SDSS data by using two galaxy cluster catalogs: one selected using the RM algorithm, and the other selected using the \citet{Yang2007} group finder. In both cases, a sharp steepening of the density profile around the clusters was observed, suggesting that the splashback feature is not purely an artifact of the RM selection. Furthermore, it was found that the splashback measurements utilizing the \citet{Yang2007} catalog agreed well with those using the RM catalog; however, the signal-to-noise of measurements using the \citet{Yang2007} catalog was not sufficient to rule out some residual systematic effect.
In addition, \citet{Baxter2017} divided the galaxy samples by color, and measured the fraction of red galaxies relative to blue galaxies as a function of cluster-centric radius.
It was found that the red fraction increased inward rapidly at approximately the measured splashback radius.
Such behavior is expected for a true physical boundary, since galaxies outside the splashback shell have never been inside the cluster and are therefore more likely to have ongoing star formation, and will thus appear bluer than galaxies which have passed through the cluster.  

\citet{Chang2017} directly investigated the potential systematic effects associated with the imposition of $R_{\lambda}$ in the RM algorithm by repeating the splashback measurements using three different richness aperture choices: 0.67, 1 and 1.5 times the original $R_{\lambda}$.
It was found that the value of $R_{\lambda}$ used to estimate richness significantly impacts the recovered splashback radius, in the same direction as suggested by \citet{Busch2017}.
While the aperture choices in that study were extreme, it suggests that the choice of the RM aperture used to estimate richness could impact the splashback radius. 

\citet{Chang2017} also used weak lensing shear estimates from the DES to measure the splashback feature around the same cluster sample, finding the location and slope of the splashback feature to be roughly consistent with that inferred from the galaxy density measurements. The splashback radius inferred from the lensing measurements was also observed to change slightly for different assumed values of $R_{\lambda}$, although the change was not significant given the low signal-to-noise of the lensing measurements.  If the splashback feature inferred from both the galaxy density and the lensing results reacts to $R_{\lambda}$ in the same way, then it is unlikely that the \citet{Busch2017} explanation above is complete. 
An alternative explanation for the observed trends is that changing $R_{\lambda}$ selects a physically different set of clusters (e.g. those that are more elongated along the line of sight), which might indeed have different splashback radii.

One way to bypass these complications is to repeat the measurements using an alternative cluster sample selected independently of the RM algorithm and of the galaxy density observations.
In this work, we measure the splashback feature around a sample of galaxy clusters identified via their Sunyaev-Zel'dovich \citep[SZ,][]{Sunyaev1972} signal in data from the South Pole Telescope (SPT) SZ survey \citep{Bleem2015} and the Atacama Cosmology Telescope Polarimeter \citep[ACTPol,][]{Hilton2017}. The SZ effect results from the cosmic microwave background photons inverse Compton scattering with hot cluster gas, and is seen as a temperature decrement at the locations of galaxy clusters in the 150 GHz maps of the SPT-SZ survey and the 148 GHz maps of the ACTPol experiment.  

Several features of the SZ-selected cluster samples used here make them useful for testing the impact of systematics on splashback measurements.
For one, the SZ observable is completely independent of all the observables in optical surveys used to measure the feature (in particular, the galaxy density).
The SZ signal is also expected to correlate more tightly with cluster mass than optical richness, reducing the impact of scatter in the mass-observable relation, therefore making it easier to compare measurements to expectations from simulations.
Additionally, SZ-selection is expected to be less affected by projection effects than optical cluster finders.
Furthermore, by selecting on the SZ signal rather than optical richness, we reduce potential correlation between the cluster selection and the quantity used to infer splashback, i.e. the galaxy density. Finally, the SZ-selected cluster samples employed here allow us to extend splashback measurements to the high-mass, high-redshift regime that has yet to be explored for splashback studies. 
We refer readers to e.g. \citet{Nagai06}, \citet{Battaglia12} and \citet{Krause12} for detailed analyses of SZ the signal-mass scaling relation.  

Finally, we note that while this work was in preparation, \citet{Zuercher2018} presented a similar analysis using clusters selected from {\it Planck} data.  Given the difference between the cluster and galaxy samples in the two works, our results can be considered complementary to theirs.

The structure of the paper is as follows.  We describe the galaxy and cluster data sets in Section~\ref{sec:data}; measurements and model fitting are described in Section~\ref{sec:measurement}; results are presented in Section~\ref{sec:result}, and we conclude in Section~\ref{sec:discussion}. Throughout this work, when calculating cosmological quantities, we use a flat $\Lambda$CDM cosmology with $H_0 = 70 \, {\rm km s^{-1} Mpc^{-1}}$, $\Omega_{\rm m} = 0.3$.
Every distance is reported in the comoving unit with $h=0.7$.

\section{Data}
\label{sec:data}

\begin{figure}
\centering
\includegraphics[width=0.99\linewidth]{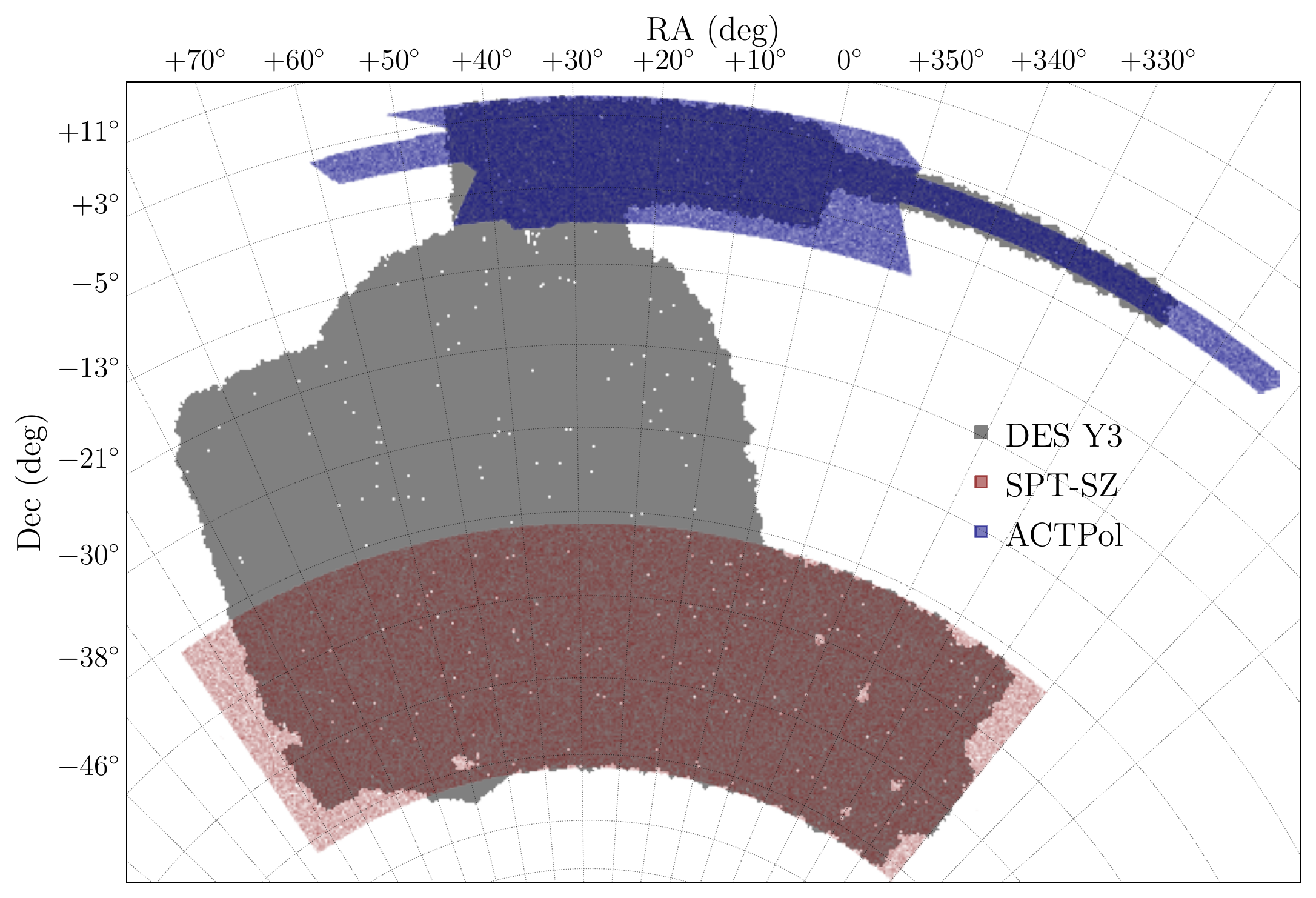}
\caption{The footprints of the DES, SPT and ACTPol. The overlapping area is $\sim$2000 ($\sim$700) ${\rm deg}^2$ between SPT (ACTPol) and DES.}
\label{fig:footprint}
\end{figure}

\subsection{SZ-selected cluster catalog from SPT}
\label{sec:sz-cl}

The SPT is a $10~{\rm m}$ millimeter/submillimeter telescope operating at the geographical South Pole \citep{Carlstrom:2011}. The cluster catalog used in this analysis was derived from data taken as part of the 2500 sq. deg. SPT-SZ survey, which mapped the sky in three frequency bands centered at 95, 150 and 220 GHz over an observation period from 2008 to 2011 \citep{Story:2013}. The construction of the catalog is described in detail in \citet{Bleem2015}. The SPT-SZ survey region is shown in Fig.~\ref{fig:footprint}.  

Clusters are identified using a linear combination of the 95 and 150 GHz SPT temperature maps adopting a matched filter approach, with the projected isothermal $\beta$-model \citep{Cavaliere1976} as the assumed source profile: 
\begin{equation}
\Delta T = \Delta T_0 (1+\theta^2/\theta_c^2)^{-1},
\end{equation}
where $\Delta T$ is the temperature in the map, and $\Delta T_0$ and $\theta_c$ are model parameters.  Filters were constructed using 12 different $\theta_c$ values between 0'.25 and 3', and applied to the maps in the Fourier domain. Cluster candidates are then identified as peaks in the filtered maps.  The maximal signal-to-noise (SNR) across these filter choices and across possible cluster positions is then considered the SNR estimate, $\xi$, for each cluster.  The sample used in this analysis uses clusters with $\xi\geq4.5$.
Follow-up optical and NIR observations are made for the 530 candidates with $\xi\geq4.7$ as well as 119 of 147 candidates down to $\xi\geq4.5$. Among these 677 candidates, 516 are confirmed by identifying an excess of clustered red-sequence galaxies and consequently given redshift and mass estimates.  Masses for each cluster are estimated using an assumed scaling relation between the SPT observable, $\xi$, and the cluster mass with a fixed $\Lambda$CDM cosmology, as described in \citet{Bleem2015}. 

Our fiducial measurements are based on a sample selected with $0.25 < z < 0.7$ and $\xi \geq 4.5$, which has 315 clusters, of which 256 are in the DES footprint.  SPT detects many clusters with $z > 1$.  However, as described in Sec.~\ref{sec:measure_sigmag}, we impose an absolute magnitude cut on the DES galaxies when correlating with the SZ-selected clusters.  Increasing the upper redshift limit of the cluster sample would necessitate using galaxies with higher luminosity to maintain completeness of the galaxy sample, thereby reducing the signal-to-noise of the splashback measurements.  Moreover, imposing the upper redshift limit enables a more direct comparison with the RM sample, which becomes increasingly incomplete beyond $z=0.7$.
The mean redshift of the selected clusters is $\langle z \rangle = 0.49$.
Adopting the mass estimates described above, the estimated mean mass of the sample is $\langle M_{\rm 500c} \rangle =3.0$$\times$$10^{14} h^{-1} M_{\odot}$.
The level of systematic uncertainties in the SPT masses in \citet{Bleem2015} is negligible for this analysis.
The masses estimated in \citet{Bleem2015}, which we use here, are obtained by assuming a fixed cosmology and running the number count experiment, yielding a mass calibration uncertainty at the 2\% level (see Section~3.1 in \citealt{Saro2015}).
On the other hand, for the same cluster sample, \citet{Bocquet18} report a mass calibration from a simultaneous fit of scaling relations, cosmology, as well as external weak-lensing data sets. 
The resultant lower and upper bounds of the mean mass with the uncertainty in cosmology from \citet{Bocquet18} are $2.5\times10^{14} h^{-1} M_{\odot}$ (-17\% relative to mean) and $3.2\times10^{14} h^{-1} M_{\odot}$ (+7\%), respectively, which we use in our analyses hereafter.
Histograms of the estimated redshifts and masses for selected clusters are shown in \Fref{fig:sample_spt}.

To reliably measure correlation functions, it is important to generate a mock cluster catalog that closely follows the survey geometry and are located at random positions.
When generating such random positions for the mock SPT catalog, we account for the non-uniformity of the cluster density across the field due to small variations in depth and apodization of the observation field boundaries.  For each field, we first generate a set of mock clusters with masses and redshifts drawn from the \citet{Tinker2008} mass function.  These mock clusters are then assigned values of the SPT observable $\xi$ using the field-dependent mass-$\xi$ relations described in \citet{Bleem2015}, applying the intrinsic and measurement scatters.  
Finally, the $\xi > 4.5$ selection is applied to the mock clusters as in the real data.\footnote{We have performed measurements with a more conservative SNR ($\xi$) cut for which every field can be assumed to be complete down to that value of $\xi$. However, using this more conservative selection does not change our results qualitatively and results in lower signal-to-noise.}

\subsection{SZ-selected cluster catalog from ACT}
\label{sec:data-act}

The ACT is a $6~{\rm m}$ telescope that is located in northern Chile \citep{ACT1,Thornton2016}. 
The ACTPol cluster sample used in this work is derived from the ACTPol two-season cluster catalog \citep{Hilton2017}. 
To extract this sample, 148\,GHz observations in a 987.5\,deg$^{2}$ equatorial field were used (\Fref{fig:footprint}), which combined data from the original ACT receiver \citep[MBAC;][]{ACT} with the first two seasons of ACTPol data.
The ACTPol survey used in this work is composed of two deep fields each of which covers $\sim$70 $ {\rm deg}^2$, taken from September 2013 to December 2013 using a single 148 GHz detector array, as well as a wider $\sim$700 $ {\rm deg}^2$ field taken from August 2014 to December 2014 with an additional 148GHz detector array \citep[see][for details on these ACTPol observations]{Naess2014, Louis2017}.
The cluster candidates were detected using a spatial matched filter based the Universal Pressure Profile \citep[UPP;][]{Nagai2007,Arnaud2010}. 
We refer readers to \citet{Hass2013} and \citet{Hilton2017} for details.
The candidates were confirmed as clusters and their redshifts measured with optical and/or IR data, mainly the Sloan Digital Sky Survey \citep[SDSS DR13;][]{Albareti2016}. Cluster masses were estimated assuming the SZ signal-mass scaling relation  and the halo mass function
from \citet{Tinker2008}, following the method in \citet{Hass2013}.
In addition, centers of the clusters are assigned as the center-of-mass of the pixels associated with the cluster that lie above the signal-to-noise ratio (SNR) of 4.
The full cluster sample in \citet{Hilton2017} are all SNR $> 4$ with mass range of roughly $1.5\times 10^{14}\,h^{-1}M_{\odot} <M_{\rm 500c,UPP}< 7\times 10^{14}\,h^{-1}$M$_{\odot}$ 
with a median mass of $M_{\rm 500c,UPP} = 2.2\times 10^{14}\,h^{-1}M_{\odot}$ and redshift range of roughly $0.15 <z< 1.4$ with a median redshift of $z = 0.49$.

\begin{figure}
\centering
\includegraphics[width=0.95\linewidth]{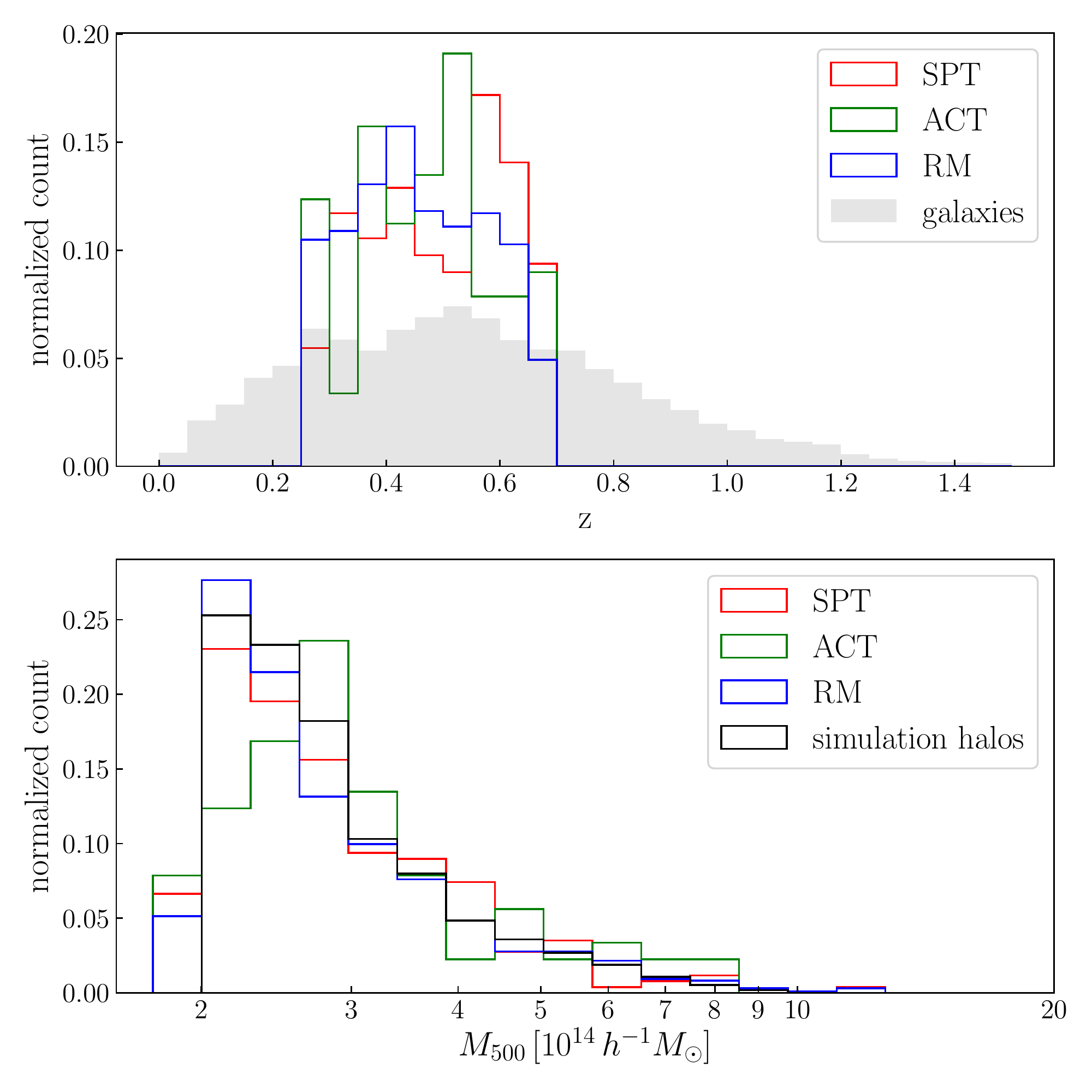}
\caption{Redshift and mass distribution for the fiducial SPT cluster sample in the DES footprint (red), the ACT cluster sample (green) and mass-matched RM clusters (blue).
Also shown are the redshift distribution of the DES Y3 galaxies (grey) in the upper panel and the mass distribution of the halos we use from the MDPL2 simulation (grey) in the lower panel.
}
\label{fig:sample_spt}
\end{figure}

In this paper we use clusters in the DES footprint and in the redshift range of [0.25,0.7].
Furthermore, the masses of the clusters are re-calibrated using the richness-mass relation from the DES Y1 analysis \citep{mcclintock2019} using clusters matched between the ACT and the DES, which gives a mass-correction factor of $1/(0.75 \pm 0.1)$.
\citet{Hilton2017} checked that the mass estimation after applying this WL-correction is consistent with that of the SPT mass (see their Fig. 25 and the associated text).\footnote{Note that \citet{Hilton2017} used a mass-richness relation from SDSS data \citep{Simet2017}, which gives a correction factor of $1/(0.68 \pm 0.11)$ that is consistent with the new value in this study.}
Applying this WL-correction with its uncertainty, there are 89 clusters with mean mass of $\langle M_{\rm 500c} \rangle =3.26^{+0.50}_{-0.39}$$\times$$10^{14} h^{-1} M_{\odot}$ and mean redshift of $\langle z \rangle = 0.49$. The redshift and mass distribution of this cluster sample is shown in Fig.~\ref{fig:sample_spt}.\footnote{Note that due to the higher noise level of the ACTPol survey than that of the SPT-SZ survey, the mean mass of ACTPol clusters is estimated higher despite the smaller SNR threshold than that of SPT-SZ.}

We generate a mock cluster catalog with random positions, which corresponds to the ACT sample in \citet{Hilton2017} by first sampling the halo mass function \citep{Tinker2008} to obtain a statistically representative sample of halos as function of mass and redshift.
Here we oversample the number of clusters in the ACT sample by a factor of 1000 to reduce the Poisson noise in the mock cluster catalog.
For each halo in the sample we calculated a filtered Compton-y signal using the matched filter and scaling relation from \citet{Hilton2017}.
Then we randomly assigned each halo a position within the ACT map footprint and compared the filtered Compton-y signal to the filter noise from \citet{Hilton2017}.
The final product is a mock halo catalog with signal-to-noise values that correspond to the filter noise in the map according to \citet{Hilton2017}.
We then apply a minimum signal-to-noise threshold of four, a redshift cut 0.7, and the signal-to-noise completeness function from \citet{Hilton2017} to account for non-uniform selection and cluster confirmation effects in the ACT sample. 

\subsection{DES Year 3 galaxy catalog}
\label{sec:data-desgold}

We measure the splashback feature around the SZ-selected clusters by correlating these clusters with galaxies, effectively using galaxies as tracers of the mass.
Our galaxy sample used for this purpose is derived from the DES data.
DES \citep{DES2005} is a five-year survey that covers $\sim5000$ square degrees of the South Galactic Cap (\Fref{fig:footprint}).
Mounted on the Cerro Tololo Inter-American Observatory (CTIO) $4~m$ Blanco telescope in Chile, the 570-megapixel Dark Energy Camera \citep{DECam} images the field in $grizY$ filters. In this analysis, we use the DES Year 3 data.\footnote{The full DES Y3 images are taken from Aug 2013 to Feb 2016.}  The raw images are processed by the DES Data Management (DESDM) system \citep{Sevilla2011,Morganson2018} and a high-quality photometric catalog (Y3 Gold v2.2) is produced after a careful subselection similar to that described in \citet{DrlicaWagner2017}. 

After filtering out stars and removing galaxies identified as failures in photometry, we apply a further magnitude and color selection with the following criteria:
$i < 22.5,\, -1 < g - r < 3,\, -1 < r - i < 2.5\,$ and $-1 < i - z < 2,$
where the color cuts are to remove color outliers that may result in catastrophic photo-z estimates \citep{crocce2018}.
We further require the error of the \textit{i}-band magnitude to be less than 0.1 to ensure good photometry and apply the DES survey depth mask (only using regions where the $i$-band magnitude limit $>$ 22.5) as well as the SPT-SZ/ACTPol survey mask, depending on the cluster catalog being used. The total number of galaxies in our sample after all these cuts is 41,102,373 (13,385,454) in the SPT (ACT) field. When performing the cluster-galaxy correlation measurements, we will also apply additional magnitude cuts as described in Sec.~\ref{sec:measure_sigmag}.

\subsection{DES Year 3 redMaPPer cluster catalog}
\label{sec:rm-dat}

The primary focus of this work is to measure the splashback radius around SZ-selected clusters. However, to further test the impact of cluster selection on splashback measurements, we also perform measurements using a catalog of optically selected clusters identified with the RM algorithm applied to the DES Y3 gold catalog (RM v6.4.22).  We apply the same redshift cut to the RM clusters as to the SPT and ACT samples.

We additionally impose a richness cut on the RM sample so that the mean mass of this sample is  matched to that of the SPT and ACT samples.  The richness cut is determined using the mass-richness relation for DES Y1 RM clusters from \citet{mcclintock2019}, which calibrated the mass through a stacked weak lensing analysis.  Using this mass-richness relation, we compute the expectation value of $M_{200m}$ for each cluster.  These are then converted into $M_{\rm 500c}$ using the mass-concentration relation of \citet{DK15} and an NFW profile as implemented in \texttt{Colossus} \citep{Diemer2017}.\footnote{http://www.benediktdiemer.com/code/colossus/}  A richness cut of $\lambda > 58$ is chosen so that the mean mass of the RM clusters is equal to that of the SPT sample.  The mean mass of the ACT sample is statistically consistent with that of the SPT sample, allowing us to use the same richness cut throughout.
The final distributions of mass and redshift are shown in \Fref{fig:sample_spt}.
In principle, uncertainty in the mass-richness relation ($\sim$5\%) could impact the mean mass of the selected RM clusters.
However, since the location of splashback scales as $M^{1/3}$, such uncertainty contributes less than  2\% uncertainty on the splashback radius, which is well below our uncertainty level of on the splashback location ($\sim$7\%).

\section{Measurement and Modeling}
\label{sec:measurement}

\subsection{Cluster profile model}
\label{sec:model}
We model the measured galaxy surface density profiles following \citet{Diemer2014} and \citet{More2015} (note that following \citealt{More2016} and \citealt{Chang2017}, we do not include a cosmic mean density term since our measurements effectively have the cosmic mean subtracted). The model comprises a ``collapsed'' part (a truncated Einasto profile) and an ``infalling'' part (a power law profile). The full model for the 3D density profile, $\rho(r)$, is
\begin{eqnarray}
\label{eq:modelstart}
\rho (r) &=& \rho^{\rm coll}(r) + \rho^{\rm infall}(r), \\[5pt]
\rho^{\rm coll}(r) &=& \rho^{\rm Ein}(r)f_{\rm trans}(r), \\[5pt]
\rho^{\rm Ein}(r) &=& \rho_{\rm s} \, {\rm exp}\Big(-\frac{2}{\alpha}\left[\left(\frac{r}{r_{\rm s}}\right)^{\alpha}-1\right]\Big), \\[5pt]
f_{\rm trans}(r) &=& \left[1+\left(\frac{r}{r_{\rm t}}\right)^{\beta}\right]^{-\gamma/\beta}, \\[5pt]
\label{eq:modelend}
\rho^{\rm infall}(r) &=& \rho_0 \left(\frac{r}{r_0}\right)^{-s_{\rm e}},
\end{eqnarray}
where we fix $r_0$ at $1.5 h^{-1} {\rm Mpc}$ and $\alpha$, $\beta$, $\gamma$, $r_{\rm s}$, $r_{\rm t}$, $\rho_{\rm s}$, $\rho_0$ and $s_{\rm e}$ are parameters of the model. Note that a transition function $f_{\rm trans}$ is needed since the splashback surface is generally not spherical in 3D so that when density is averaged in a spherical shell, the density drop gets smeared out (see \citealt{mansfield2017} for details).
Also, the model is flexible enough to provide profiles that are featureless, as can be seen in \Fref{fig:prof_color} (see also \citet{Baxter2017} for an extensive model comparison).
We have also experimented with truncating the infalling power-law term at small radii via $\rho^{\rm infall}(r) = \rho_0 \big[1/\rho_{\rm max} + (r/r_0)^{s_{\rm e}}\big]^{-1}$, where $\rho_{\rm max}$ sets the maximum density of the infalling term at small scales \citep{Diemer2017}.
However, implementing such truncation has a negligible impact, and so for simplicity we leave it out.

The splashback radius, $r_{\rm sp}$, is a derived parameter in this
model, and represents the minimum of the logarithmic derivative ${\rm d}\log \rho/ {\rm d}\log r$ of the total density.

Below, we will measure the projected galaxy surface number density around clusters. We relate the 2D projected density to the 3D density via
\begin{equation}
\Sigma(R) = \int^{l_{\rm max}}_{-l_{\rm max}} {\rm d} l \, \rho \left(\sqrt{R^2 + l^2} \right),
\end{equation}
where $R$ is the 2D projected distance from the cluster center and $l_{\rm max}$ is the maximum line-of-sight distance of integration, set to $40\,h^{-1}{\rm Mpc}$. We have checked that extending $l_{\rm max}$ to a higher value does not change our result significantly.  

To account for the effects of cluster mis-centering, which can be different for the SZ-selected clusters and for the RM-selected clusters, we assume that some fraction, $f_{\rm mis}$, of the clusters are miscentered, while a fraction $(1-f_{\rm mis})$ are correctly centered.
The observed profile is then
\begin{equation}
\label{eq:miscenstart}
\Sigma = (1-f_{\rm mis})\Sigma_{\rm 0} + f_{\rm mis}\Sigma_{\rm mis},
\end{equation}
where $\Sigma_0$ is the profile without mis-centering and $\Sigma_{\rm mis}$ is the average density profile of the mis-centered clusters. 

For a cluster miscentered by a distance $R_{\rm mis}$, the azimuthally averaged profile is		
\begin{equation}
\Sigma_{\rm mis}(R|R_{\rm mis}) = \\
\int^{2\pi}_0 \frac{{\rm d}\theta}{2\pi}\Sigma_0\Big(\sqrt{R^2 + R^2_{\rm mis} + 2RR_{\rm mis}{\rm cos}\theta}\Big).
\end{equation}
The profile averaged over a distribution of $R_{\rm mis}$ is then
\begin{equation}
\Sigma_{\rm mis} (R) = \int {\rm d}R_{\rm mis}P(R_{\rm mis})\Sigma_{\rm mis}(R|R_{\rm mis}),
\end{equation}
where $P(R_{\rm mis})$ is the probability distribution of a cluster to be mis-centered by a distance $R_{\rm mis}$ from the true center. Note that this model is not sample-specific, but rather can be applied to any cluster sample.

We assume the miscentering distribution is described by a two-dimensional Gaussian.
In this case, the distribution of $R_{\rm mis}$ can be characterized with a Rayleigh distribution:
\begin{equation}
\label{eq:miscenend}
P(R_{\rm mis}) = \frac{R_{\rm mis}}{\sigma^2_R} {\rm exp}\left[-\frac{R^2_{\rm mis}}{2\sigma^2_R}\right].
\end{equation}
\citet{Saro2015} and \citet{Rykoff2016} show that the Rayleigh distribution can be used to describe the miscentering of SPT and RM clusters, respectively.

For the SPT-selected sample, the positional uncertainty, which depends on the SPT beam size and the cluster size, applies to all clusters. For these clusters we therefore set $f_{\rm mis} = 1$. \citet{Saro2015} also provide the positional uncertainty of the individual SPT clusters in units of arcminutes (see their Eq. 11), which we convert into a distance unit, taking into account the redshift values of the clusters. The calculated positional uncertainty of the SPT sample is $\ln (\sigma_{\rm R}/{(h^{-1}\rm Mpc\,})) = -2.7 \pm 0.4$, which we use as the prior for the SPT miscentering.  Note that we assume the positional uncertainty of the measured galaxy surface density profile is constant over the entire cluster sample, which is not completely accurate.  However, the validity of this approach in cluster density profile measurements is broadly confirmed by previous studies, e.g. \citet{Baxter2017,Chang2017,mcclintock2019}.

On the other hand, for the RM clusters, the centering distribution is expected to be bimodal, with some clusters being perfectly centered and some being miscentered. We therefore apply priors of $f_{\rm mis} = 0.22 \pm 0.11$ and $\ln (\sigma_{\rm R}/{(h^{-1}\rm Mpc\,})) = -1.19 \pm 0.22$ following \citet{Rykoff2016}.

The miscentering distribution of ACT clusters differs from that of SPT because of the different beam size and because we use a lower signal-to-noise threshold when constructing the ACT sample. Given the large increase in cluster miscentering at low signal-to-noise, we opt to use RM-derived centers for the ACT clusters when possible.  
Out of 89 ACT clusters in the DES footprint, 80 of them are matched to RM clusters. 
For the remaining 9 clusters, we use the locations of the cluster BCGs as the center, where the BCG is identified by inspection. We note that for the 80 ACT clusters that are matched to RM clusters, their BCG positions typically agree precisely with those of the corresponding RM centers (66 out of 80).  Since most of the ACT clusters are assigned RM center, we adopt the RM miscentering priors for describing the miscentering of the ACT clusters.

%%%%%%%%%%%%%%%%%%%%%%%%%%
\begin{table}
	\centering
	\begin{tabular}{lll}
		Parameter & Prior & description \\ \hline
		$\log \rho_{\rm s}$ & $[-\infty,\infty]$ &  amplitude of the Einasto profile \\
        $\log \alpha$ & $\mathcal{N}(\log(0.22),0.6^2)$ & parameter of the Einasto profile\\
        $\log r_{\rm s}$ & $[\log(0.1),\log(5.0)]$ & scale radius of the Einasto profile\\
		$\log r_{\rm t}$ & $[\log(0.5),\log(5.0)]$ & scale radius of $f_{\rm trans}$\\
		$\log \beta$ & $\mathcal{N}(\log(6.0),0.2^2)$ & first slope parameter of $f_{\rm trans}$\\
		$\log \gamma$ & $\mathcal{N}(\log(4.0),0.2^2)$ & second slope parameter of $f_{\rm trans}$\\
		$\log \rho_{\rm 0}$ & $[-\infty,\infty]$ &  amplitude of $\rho_{\rm infall}$\\
        $s_{\rm e}$ & [0.1,10.0] & log-slope of $\rho_{\rm infall}$\\
        $\ln \sigma_{\rm R}$ & $\mathcal{N}(-2.7,0.4^2)$ (SPT) & miscentering amplitude\\
        & $\mathcal{N}(-1.19,0.22^2)$ (RM/ACT)\\ 
        $f_{\rm mis}$ & $1.0$ (SPT) & miscentering fraction\\
        & $\mathcal{N}(0.22,0.11^2)$ (RM/ACT)
	\end{tabular}
    \caption{Prior range of each model parameter. $\mathcal{N}(m,\sigma^2)$ represents a Gaussian prior with  mean $m$ and  standard deviation $\sigma$ (see Sec.~\ref{sec:model} and ~\ref{sec:model-fitting})}.
\label{tab:modeling_parameters}
\end{table}
%%%%%%%%%%%%%%%%%%%%%%%%%%%%

\subsection{Measurement of the galaxy surface density profile}
\label{sec:measure_sigmag}

We adopt the same method for measurement of the galaxy surface density profile, $\Sigma_{g}$, as in \citet{Baxter2017} and \citet{Chang2017}. The mean galaxy distribution around clusters can be related to the cluster-galaxy cross-correlation function, $\omega(R)$, where $R$ represents the 2D projected {\it comoving} distance from the cluster center. As shown in \citet{Diemer2014}, $r_{\rm sp}$ is expected to scale with {\it physical} $R_{\rm 200m}$ on average. Therefore, since physical $R_{\rm 200m}$ is proportional to $(1+z)^{-1}$ for a fixed mass, measuring $\omega(R)$ in comoving units, $R_{\rm com} = (1+z)R_{\rm phys}$, automatically accounts for this redshift dependence of $r_{\rm sp}$.

We divide the cluster sample into redshift bins with $\Delta z = 0.025$.
Then we measure the mean cluster-galaxy angular correlation function in the i-th bin, $\omega(\theta,z_i)$, using the Landy-Szalay estimator \citep{Landy1993}. $\omega(\theta,z_i)$ is converted to $\omega(R,z_i)$ assuming the midpoint redshift value of the redshift bin.\footnote{We have checked that this approximation introduces a negligible impact on the measured correlation function, with respect to the level of uncertainty.} Next we average $\omega(R,z_i)$ into the mean $\omega(R)$, weighted by the number of clusters in each redshift bin. 
 
Finally, the measured correlation function is related to the mean-subtracted density profile, $\Sigma_g(R)$, via
\begin{equation}
\Sigma_g(R) = \bar{\Sigma}_g \omega(R),
\end{equation}
where $\bar{\Sigma}_g$ is the mean surface density of galaxies averaged over redshift bins, weighted by the number of clusters in each bin.

We apply an approximate absolute magnitude cut on the galaxies following the method of \citet{More2016}. That is, for each redshift bin, we apply an absolute magnitude cut corresponding to the apparent magnitude cut ($i<22.5$) at the maximum redshift of the cluster sample, 0.7. When calculating the absolute magnitudes of galaxies, we assume all the galaxies have the same redshift as the cluster of interest. For our sample and redshift range, this luminosity cut is $M^{*}_i \equiv M_i - 5{\rm log}(h) < -19.87$. After applying this absolute magnitude cut, the total number of galaxies ranges from 4,780,059 for the lowest redshift bin to 39,117,782 for the highest redshift bin.\footnote{Note that we do not use photometric redshift information of galaxies in this method. Rather, the correlation function automatically picks up galaxies that are correlated with the clusters.}

The covariance matrix of the measurements is constructed using jackknife resampling \citep[e.g.,][]{Norberg:2009}. For this purpose, we divide the survey area into 100 approximately equal area subregions. Each subregion is approximately $4.4\times4.4$ square degrees, significantly larger than the maximum scales considered in this analysis.

%%%%%%%%%%%%%%%%%%%%%%%%%%
\setlength{\tabcolsep}{5pt}

\begin{table*}
	\centering
    \bgroup
    \def\arraystretch{1.0}
	\begin{tabular}{cccccccccccc}
		Sample & $\log \alpha$ & $\log r_{\rm s}$ & $\log r_{\rm t}$ & $\log \beta$ & $\log \gamma$ & $s_{\rm e}$ & $f_{\rm mis}$ & $\ln \sigma_{\rm R}$ & $r_{\rm sp}$ [$h^{-1}$Mpc] & $\frac{d\log\rho}{d\log r}(r_{\rm sp})$ & $\frac{d\log\rho_{\rm coll}}{d\log r}(r_{\rm sp})$ \\ \hline
		SPT & $-0.92^{+0.22}_{-0.44}$ & $-0.61^{+0.26}_{-0.18}$ & $0.34^{+0.14}_{-0.12}$ & $0.78^{+0.15}_{-0.25}$ & $0.60^{+0.17}_{-0.23}$ & $1.66^{+0.38}_{-0.47}$ & $1.0$ & $-2.00^{+0.01}_{-1.01}$ & $2.37^{+0.51}_{-0.48}$ & $-3.47^{+0.43}_{-0.30}$ & $-5.17^{+1.06}_{-0.60}$\\
        ACT & $-0.88^{+0.27}_{-0.32}$ & $-0.77^{+0.38}_{-0.09}$ & $0.30^{+0.19}_{-0.15}$ & $0.80^{+0.13}_{-0.29}$ & $0.60^{+0.17}_{-0.24}$ & $1.28^{+0.68}_{-0.82}$ & $0.20^{+0.10}_{-0.09}$ & $-1.19^{+0.21}_{-0.24}$ & $2.22^{+0.72}_{-0.56}$ & $-3.92^{+0.86}_{-0.51}$ & $-5.40^{+1.27}_{-0.58}$\\
		DES & $-1.16^{+0.18}_{-0.46}$ & $-0.67^{+0.28}_{-0.20}$ & $0.22^{+0.06}_{-0.05}$ & $0.88^{+0.11}_{-0.18}$ & $0.65^{+0.16}_{-0.17}$ & $1.69^{+0.09}_{-0.15}$ & $0.12^{+0.07}_{-0.06}$ & $-1.15^{+0.22}_{-0.31}$ & $1.88^{+0.13}_{-0.12}$ & $-3.71^{+0.30}_{-0.20}$ & $-5.52^{+0.88}_{-0.61}$ \\
        SPT red & $-0.73^{+0.08}_{-0.28}$ & $-0.63^{+0.10}_{-0.23}$ & $0.39^{+0.14}_{-0.10}$ & $0.81^{+0.14}_{-0.26}$ & $0.60^{+0.16}_{-0.24}$ & $1.44^{+0.19}_{-0.64}$ & $1.0$ & $-2.68^{+0.50}_{-0.40}$ & $2.64^{+0.57}_{-0.34}$ & $-4.05^{+0.48}_{-0.39}$ & $-5.63^{+1.19}_{-0.52}$ \\
        SPT green & $-0.66^{+0.26}_{-0.48}$ & $0.03^{+0.43}_{-0.15}$ & $0.26^{+0.17}_{-0.09}$ & $0.77^{+0.20}_{-0.19}$ & $0.58^{+0.18}_{-0.22}$ & $1.50^{+0.30}_{-0.78}$ & $1.0$ & $-2.68^{+0.42}_{-0.41}$ & $2.16^{+0.71}_{-0.27}$ & $-3.73^{+0.50}_{-0.62}$ & $-5.11^{+0.96}_{-0.92}$ \\
        DES red & $-1.07^{+0.20}_{-0.06}$ & $-0.95^{+0.30}_{-0.01}$ & $0.25^{+0.06}_{-0.03}$ & $0.91^{+0.10}_{-0.17}$ & $0.70^{+0.15}_{-0.18}$ & $1.68^{+0.06}_{-0.15}$ & $0.09^{+0.07}_{-0.05}$ & $-1.14^{+0.22}_{-0.35}$ & $2.02^{+0.12}_{-0.09}$ & $-4.13^{+0.31}_{-0.23}$ & $-6.00^{+0.87}_{-0.71}$\\
        DES green & $-0.73^{+0.34}_{-0.13}$ & $0.18^{+0.03}_{-0.24}$ & $0.18^{+0.09}_{-0.02}$ & $0.90^{+0.14}_{-0.19}$ & $0.64^{+0.19}_{-0.15}$ & $1.63^{+0.14}_{-0.13}$ & $0.24^{+0.10}_{-0.11}$ & $-1.17^{+0.26}_{-0.21}$ & $1.81^{+0.13}_{-0.14}$ & $-3.75^{+0.24}_{-0.60}$ & $-5.53^{+0.48}_{-1.50}$
	\end{tabular}
    \egroup
    \caption{1$\sigma$ ranges of the best-fit parameters in different samples, including the model parameters (Sec.~\ref{sec:model}), splashback location ($r_{\rm sp}$) and the minimum logarithmic slope at $r_{\rm sp}$. We also show the 1$\sigma$ range of the logarithmic derivative of $\rho_{\rm coll}$.
The values of mean mass (redshift) of the SPT, ACT and RM samples are $M_{\rm 500c} = 3.0\times10^{14}h^{-1}M_{\odot}$ (0.49), $M_{\rm 500c} = 3.3\times10^{14}h^{-1}M_{\odot}$ (0.49) and $M_{\rm 500c} = 3.0\times10^{14}h^{-1}M_{\odot}$ (0.46), respectively. 
Note that we do not show results of $\rho_0$ and $\rho_{\rm s}$, since they do not contain much physical information determining $r_{\rm sp}$. `Red' and 'green' represent the galaxy colors as defined in Sec.~\ref{sec:color_def}.}
    \label{tab:results}
\end{table*}
%%%%%%%%%%%%%%%%%%%%%%%%%%%%

\subsection{Model fitting}
\label{sec:model-fitting}

Given the jackknife estimate of the covariance matrix, $\textbf{C}$, we adopt a Gaussian likelihood, $\mathcal{L}$, for the data, $\vec{d}$, given the model parameters, $\vec{\theta}$:
\begin{equation}
\ln \mathcal{L}[\vec{d} | \vec{m}(\vec{\theta}) ] = -\frac{1}{2}\left[\vec{d} - \vec{m}(\vec{\theta})\right]^{\rm T} \textbf{C}^{-1} \left[\vec{d} - \vec{m}(\vec{\theta}) \right],
\end{equation} 
where $\vec{m}(\vec{\theta})$ is the model evaluated at the parameter values specified by $\vec{\theta}$. The posterior on the model parameters is then given by 
\begin{equation}
\ln \mathcal{P}(\vec{\theta} | \vec{d}) = \ln \big[ \mathcal{L}(\vec{d} | \vec{m}(\vec{\theta}) ){\rm Pr}(\vec{\theta}) \big],
\end{equation}
where ${\rm Pr}(\vec{\theta})$ are the priors imposed on $\vec{\theta}$.

We draw samples from the posterior on the model parameters using Markov Chain Monte Carlo method of \citet{MCMC} as implemented in the code \texttt{emcee} \citep{emcee2013}. We consider eight free parameters ($\rho_0$, $\rho_s$, $r_{\rm t}$, $r_{\rm s}$, $\alpha$, $\beta$, $\gamma$ and $s_{\rm e}$) from the cluster profile model and one parameter ($\ln \sigma_{\rm R}$) from the miscentering model (Eqs.~\ref{eq:miscenstart}-\ref{eq:miscenend}).
While the number of free parameters is large relative to the number of data points (12 for RM clusters, 9 for SZ clusters), our main intention here is not to extract robust constraints on the model parameters, but rather to use the model fits to smoothly interpolate the data to extract constraints on its logarithmic derivative.

We adopt priors similar to those used by \citet{Chang2017}, with a modification in the prior in the Einasto slope parameter, $\alpha$, since $\alpha$ is known to be dependent on the halo mass \citep{Gao2008}.
The details, including the adopted priors, are summarized in Table~\ref{tab:modeling_parameters}.
Note that when we fit the RM and ACT cluster profiles, we vary $f_{\rm mis}$, adopting the prior of \citet{Rykoff2016}.
Thus, there is one additional free parameter in that case.

The model introduced above is not expected to be a good fit beyond about $9 R_{\rm vir}$ \citep{Diemer2014}, where a simple power law model no longer holds for the infall term. Hence, we restrict the range of $R$ to $0.2-10 h^{-1}{\rm Mpc}$. We exclude the scale below $0.2 h^{-1} {\rm Mpc}$ since the crowdedness of cluster fields and the existence of the BCG make the galaxy density measurements in this regime somewhat suspect.

\section{Results}
\label{sec:result}

\subsection{Splashback feature around SZ-selected clusters}
\label{sec:act}

\begin{figure}
\centering
\includegraphics[width=1.0\linewidth]{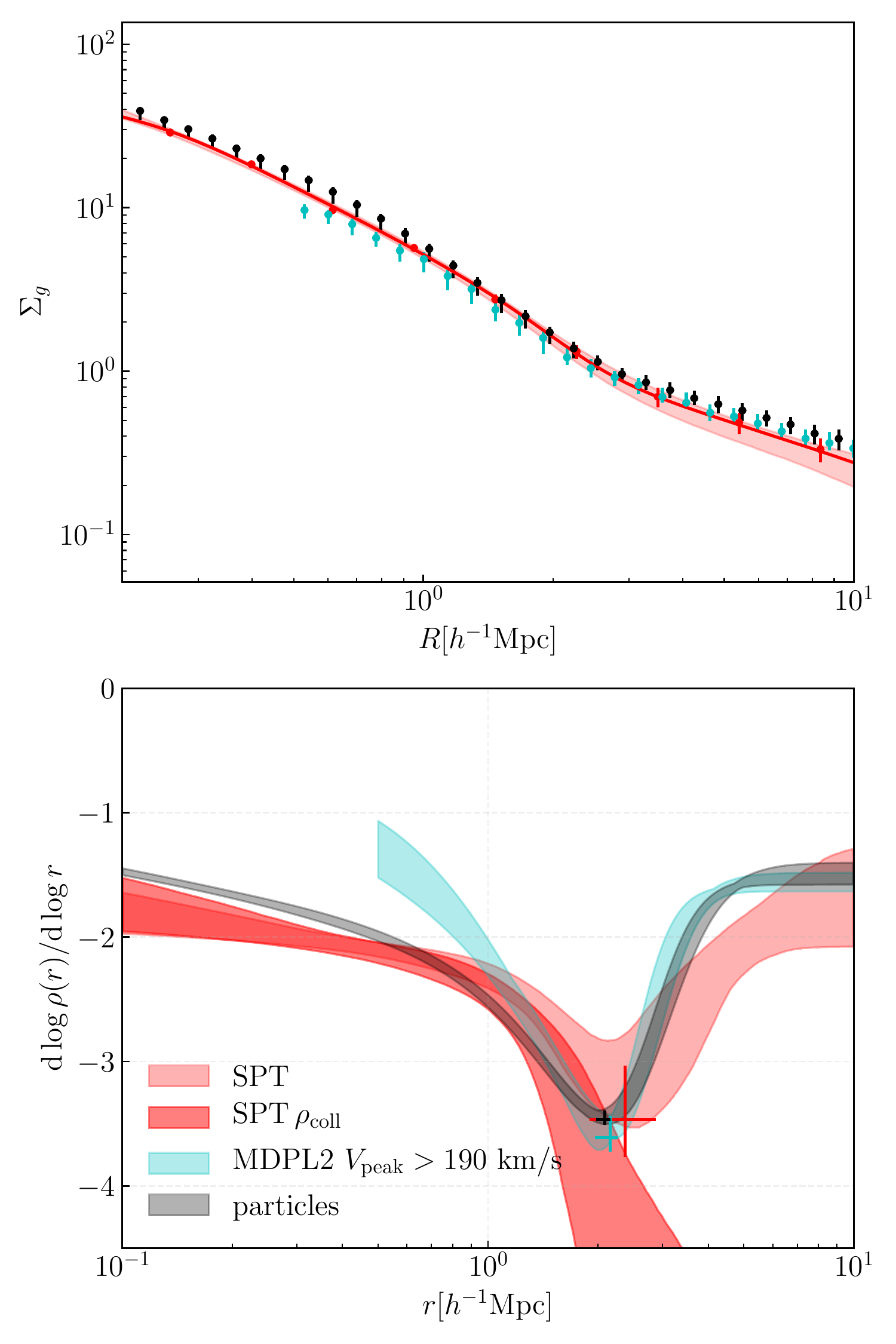}
\caption{The mean-subtracted 2D galaxy density profile, $\Sigma_g$,  around SPT SZ-selected clusters (top) and  logarithmic derivatives of the model fit 3D density profile (bottom).
The band in light red in the top panel represents the 1$\sigma$ range of the fitted profile.
Also shown are the profiles and logarithmic derivative profiles from the measurements in simulations (subhalos: cyan, particles: black).
Note that the profiles for the particles are re-normalized for an easier comparison.
The bands in the bottom panel represent the 1$\sigma$ range of the logarithmic derivative of the total density profile, $\rho(r)$, while the band in dark red corresponds to the profile of the collapsed term, $\rho^{\rm coll}(r)$, alone.
The 1$\sigma$ ranges for $r_{\rm sp}$ and the corresponding profile slope are shown with crosses with the corresponding colors.
The uncertainties for the simulation profiles include the cosmology uncertainty for SPT cluster masses (see Sec.~\ref{sec:sz-cl}).
}
\label{fig:sigmag_sptsim}
\end{figure}

In the upper panel of \Fref{fig:sigmag_sptsim}, we show the galaxy density profiles measured around the SPT SZ-selected clusters and the best-fit model profile in red. 
In the lower panel, we show the 68\% confidence interval on the logarithmic derivative of the 3D galaxy density profile, $\rho(r)$, inferred from the model fits (light red band).
Also shown is the inferred logarithmic derivative of $\rho^{\rm coll}(r)$ (the inner collapsed profile) inferred from the same joint model fits (dark red band).
$r_{\rm sp}$ is defined as the minimum of the logarithmic derivative of the total density profile, $\rho(r)$.
We report the constraints on all the model parameters in Table~\ref{tab:results}. 

The inferred logarithmic derivative profile exhibits a steepening at $\sim$2 Mpc, which we identify as the splashback feature.
The best-fit $r_{\rm sp}$ and its uncertainty for the SPT sample is $2.37^{+0.51}_{-0.48}$ $h^{-1}$Mpc and the logarithmic slope at $r_{\rm sp}$ is $-3.47^{+0.43}_{-0.30}$. The inferred logarithmic slope of $\rho^{\rm coll}$ at $r_{\rm sp}$ is $-5.17^{+1.06}_{-0.60}$, significantly steeper than the maximum logarithmic slope obtained by an NFW profile ($-3$).  As we discuss in the next section, the slope of the total profile appears to be consistent with expectations from N-body simulations.

\begin{figure}
\centering
\includegraphics[width=1.0\linewidth]{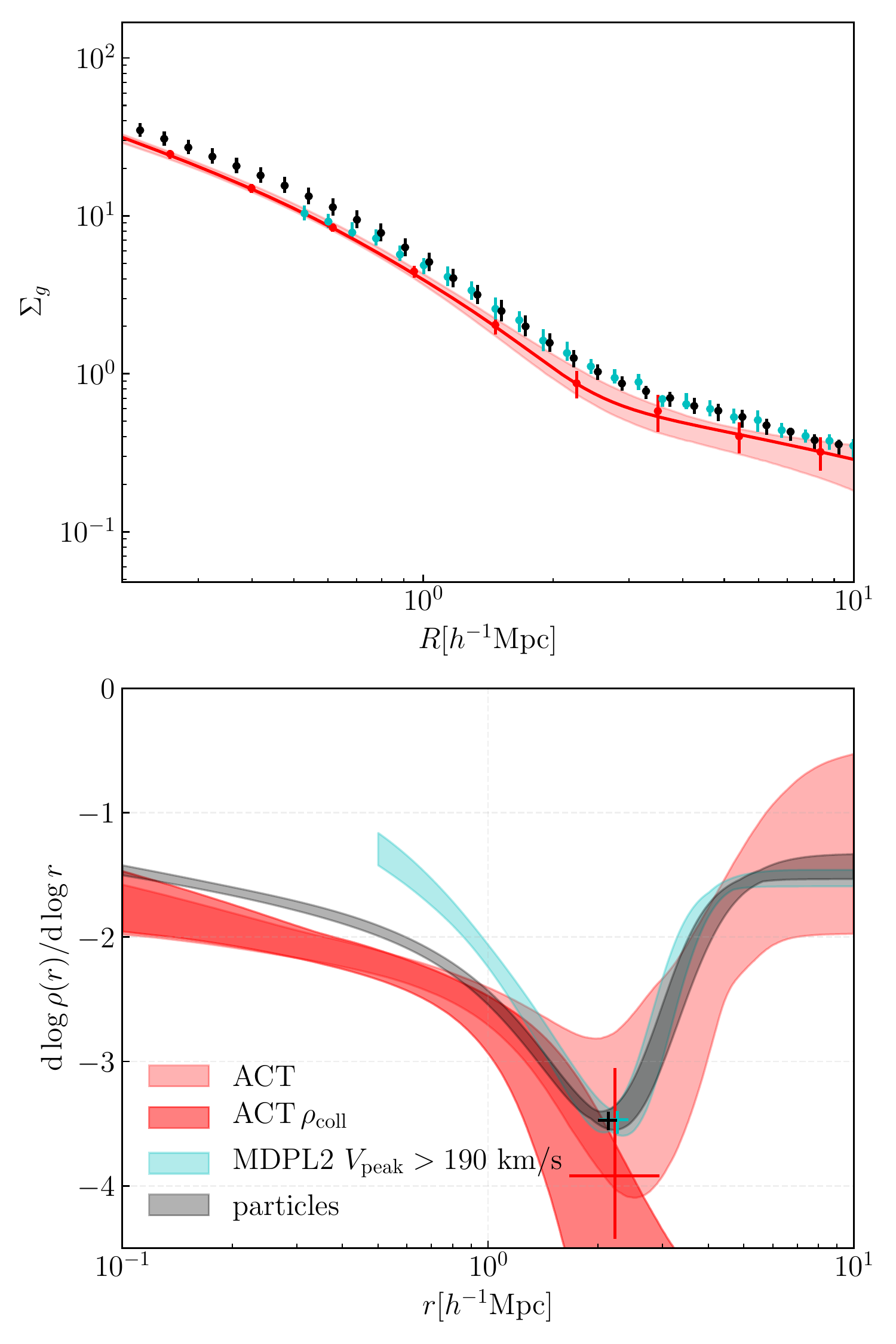}
\caption{Same as \Fref{fig:sigmag_sptsim}, but using ACT-selected clusters. The uncertainties for the simulation profiles include the WL-calibration uncertainty for ACT cluster masses (see Sec.~\ref{sec:data-act}). 
}
\label{fig:sigmag_act}
\end{figure}

The analogous plot for the SZ-selected clusters from ACT is shown in Fig.~\ref{fig:sigmag_act}.
Again, we find evidence for a steepening of the logarithmic derivative of $\rho(r)$ at roughly 2 Mpc, consistent with expectation from N-body simulations.  We measure the ACT-selected clusters to have $r_{\rm sp} = 2.22^{+0.72}_{-0.56}$ $h^{-1}$Mpc with the steepest slope of $-3.92^{+0.86}_{-0.51}$.
The inferred logarithmic slope of $\rho^{\rm coll}$ at $r_{\rm sp}$ is $-5.40^{+1.27}_{-0.58}$, significantly steeper than the steepest slope obtained by an NFW profile ($-3$).

The measurements of the galaxy density profiles around the SPT and ACT-selected clusters, taken together, constitute strong evidence for  detection of a splashback feature around SZ-selected clusters.  For both samples, the inferred logarithmic slope of the collapsed profile ($\rho_{\rm coll}$) at $r_{\rm sp}$ is steeper by $\sim 2\sigma$ than the minimum slope attained $-3$ by an NFW profile.  These findings add significant weight to claims that splashback has been detected in the galaxy density profiles around massive clusters.  The SZ-selected clusters are not as sensitive to many of the RM-related selection effects that could potentially mimic a splashback feature discussed in \citet{Busch2017} and \citet{Chang2017}.

\subsection{Comparison with simulations}
\label{sec:comp-sim}

We now compare our measurement of the splashback feature to predictions from
cosmological dark matter-only N-body simulations. Rather than attempting to populate these simulations with galaxies, we instead compare the measurements to both subhalos and particles from the simulations.  The simulated profiles of subhalos and particles are derived from the publicly available Multidark catalogs \citep{Multidark}.\footnote{https://www.cosmosim.org/}  These simulations use a 1$h^{-1}$Gpc box, with $3840^3$ particles with a mass resolution of $1.5 \times 10^9 M_\odot h^{-1}$ and $\Omega_{\rm m}=0.307$ and $h=0.677$.
We match the SZ cluster selection of SPT and ACT to simulations by adopting a lower mass threshold such that the mean mass of our sample matches that of the observed sample.
\footnote{We also have generated profiles of the simulated halos with the SZ selection from the SPT/ACT by adopting the mass-observable relations in \citet{Bocquet18} and \citet{Hilton2017} with the intrinsic and measurement scatters in the relations and applying the same SNR cuts as to the data. We checked that it negligibly affects the location and the depth of the splashback feature, compared with the uncertainty level of our data.}
We use the MultiDark Planck 2 (MDPL2) snapshot at the redshift of $z=0.49$, which is the closest snapshot to the observed mean redshift of our sample available for both particles and subhalos. 
We have checked that using halos from all snapshots between $z=0.25$ and $z=0.7$ does not significantly change the location of splashback feature.
We measure the splashback radius from the 3D density profile of the selected halos.

In \Fref{fig:sigmag_sptsim}, we show the comparison of the logarithmic slope of the number density of subhalos (cyan) and DM particles (black) in simulations to the slope of the number density profile of galaxies in SPT clusters.
As described above, the halo sample identified in the simulations has been chosen to have the same mean mass as the SPT-selected clusters, $3.0^{+0.2}_{-0.5}\times 10^{14} h^{-1} M_\odot$. The subhalo curve is based on subhalos with $V_{\rm peak} > 190$ km/s which was chosen to roughly match the amplitude of the number density profile of galaxies around these clusters.  The minimum of the slope for the simulation is at $2.16^{+0.10}_{-0.20}$ $h^{-1}$Mpc (subhalos) and $2.08^{+0.08}_{-0.11}$ $h^{-1}$Mpc (particles).  The observed splashback radius in the data is $2.37^{+0.51}_{-0.48}$ $h^{-1}$Mpc, which is in agreement with the simulations within $1\sigma$.

In \Fref{fig:sigmag_act}, we show a comparison of the splashback feature measured with the ACT sample and the corresponding simulation profiles from subhalos and particles.
The halos that are mass-matched to the ACT clusters  have mean mass of $M_{\rm 500c}=3.3^{+0.5}_{-0.4}$$\times$$h^{-1} 10^{14}M_{\odot}$.\footnote{The mass uncertainty of the ACT clusters comes from the uncertainty of the weak-lensing mass calibration applied in \citet{Hilton2017} (see Sec.~\ref{sec:data-act} for details). This uncertainty in mass is reflected in the error bars of the simulation surface density profiles in \Fref{fig:sigmag_act}.}
We measure the splashback radius for the ACT clusters to be $2.22^{+0.72}_{-0.56}$$h^{-1}\,{\rm Mpc}$.  This is consistent within $1\sigma$ of the splashback radius measured for mass-matched halos in the simulation: $2.26^{+0.15}_{-0.25}$$h^{-1}$Mpc for subhalos and $2.13^{+0.12}_{-0.14}$$h^{-1}$Mpc for particles.

In \Fref{fig:sigmag_sptsim} and \Fref{fig:sigmag_act}, at small radii, the slope of the subhalo profile in simulations is much shallower than that of the observed galaxy profiles due to the disruption of subhalos in the simulations. The subhalos lose mass due to tidal interactions and pass below the resolution limit in the central regions, resulting in a flattening of the inferred slope. Dark matter particles are expected to trace the galaxies more closely in the inner regions than subhalos, as the visible parts of galaxies are not disrupted completely by tidal stripping. 

However, the observed galaxy profiles of SPT and ACT are noticeably steeper in the inner regions (radii smaller than $\sim0.5h^{-1}{\rm Mpc}$) than that from simulation DM particles. While this has been previously noted by \citet{masjedi2006} and \citet{watson2010} for Luminous Red Galaxies (LRGs), our results show that this difference also exists around massive clusters using galaxy samples down to lower mass.
Models of galaxy evolution show \citep[e.g., see Fig. 10 in][and associated discussion]{budzynski12} that steepness of the radial profile of the galaxy number density
is sensitive to the model assumptions about survival of the stellar component of galaxies against tidal disruption and treatment of dynamical friction. Detailed comparisons
of the matter and galaxy density profiles of the kind shown here and interpretation of the differences is outside the scope of this paper, but we note that such comparisons
along with the interpretation of the trends of the splashback radius discussed below, will provide useful constraints on quenching processes and their time scales, as well as 
on details of dynamical processes and time scales of galaxy disruption and merging due to tidal forces and dynamical friction.

\begin{figure}
\centering
\includegraphics[width=1.0\linewidth]{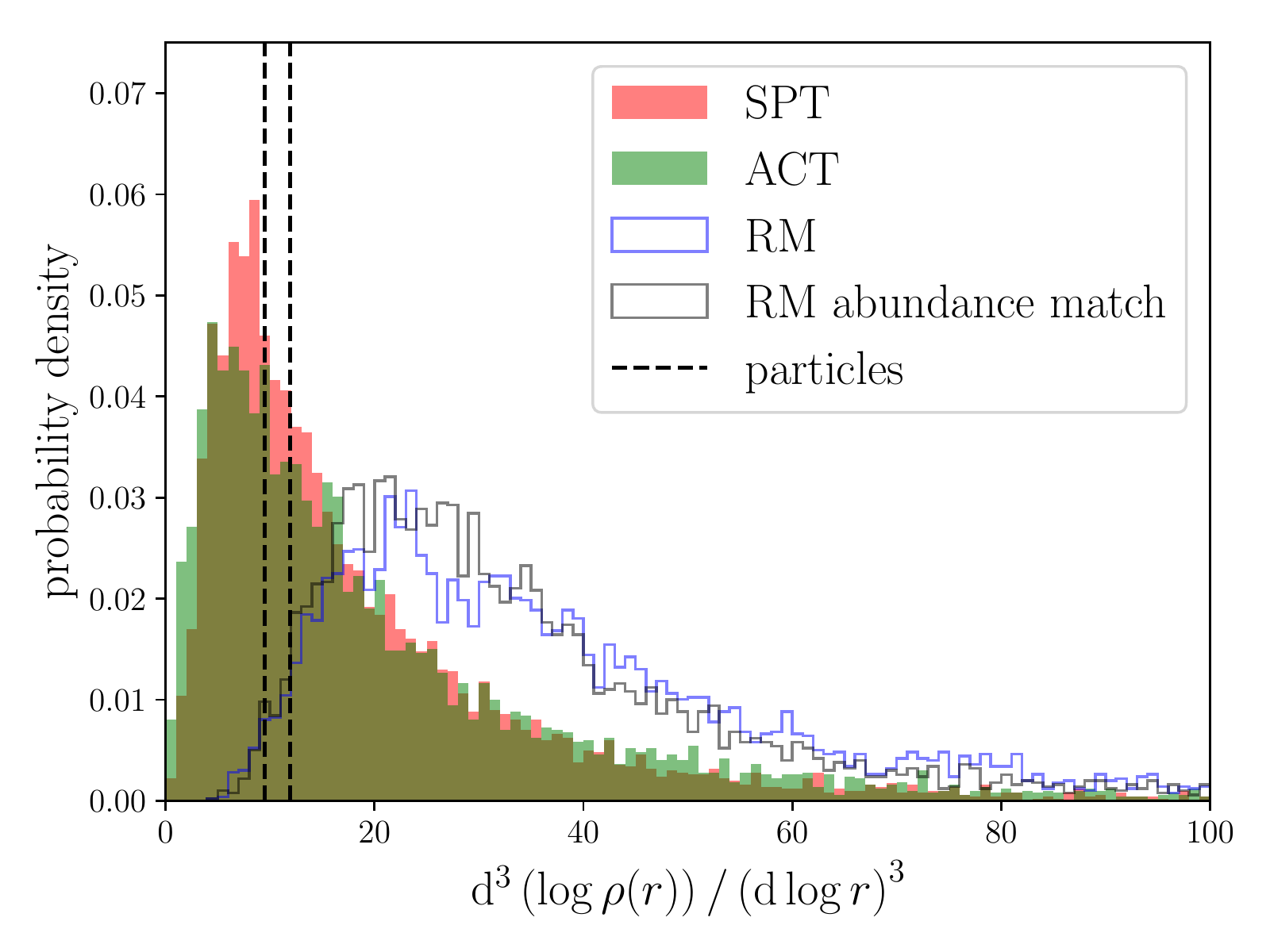}
\caption{The posterior distributions of the third logarithmic derivatives of the fitted 3D density profiles evaluated at the splashback radius (red: SPT; green: ACT, blue: RM). This quantity represents the curvature at the location of splashback or in other words the width of the splashback feature. 
The dashed black lines represent the 1$\sigma$ range for the particle profile. It is evident that the SZ selected clusters are consistent with simulations while RM clusters have a narrower splashback feature. 
}
\label{fig:2nd-deriv}
\end{figure}

One way to compare the measured and simulated density profiles is to examine the third derivatives of these profiles at the splashback radius. Since the splashback feature appears as a narrow minimum in the logarithmic derivative of the profile, the third derivative of the profile at splashback effectively measures the width of this minimum. This comparison is shown in \Fref{fig:2nd-deriv}. We see that both the SPT and ACT-selected cluster samples exhibit consistent third logarithmic derivatives at splashback (1-$\sigma$ ranges of [6.1, 31.7] for SPT and [5.3, 38.4] for ACT), and that these measurements are consistent with expectations from simulations ([9.5, 11.9]).

In summary, we find that for the SZ-selected samples the measured splashback radii are statistically consistent with expectations from simulations. This is in contrast to previous measurements of galaxy profiles around RM clusters, for which the splashback radii were inferred to be significantly smaller than predicted from simulations. We note, though, that the statistical uncertainties of our measurements with SZ-selected clusters are larger than previous RM cluster measurements because the SZ clusters have a higher mass threshold, therefore a smaller sample size.

The full profiles of the SZ-selected clusters also appear to be similar to expectations from simulations, as seen in \Fref{fig:sigmag_sptsim}, \Fref{fig:sigmag_act}, and \Fref{fig:2nd-deriv}, except in the central regions, where surface density profiles of galaxies appear to be steeper than that of particles in the simulation. The apparent consistency of the SZ cluster measurements and the subhalo measurements from simulations is confirmed via a $\chi^2$ test (Sec.~\ref{sec:comparison_profile}).

\begin{figure}
\centering
\includegraphics[width=1.0\linewidth]{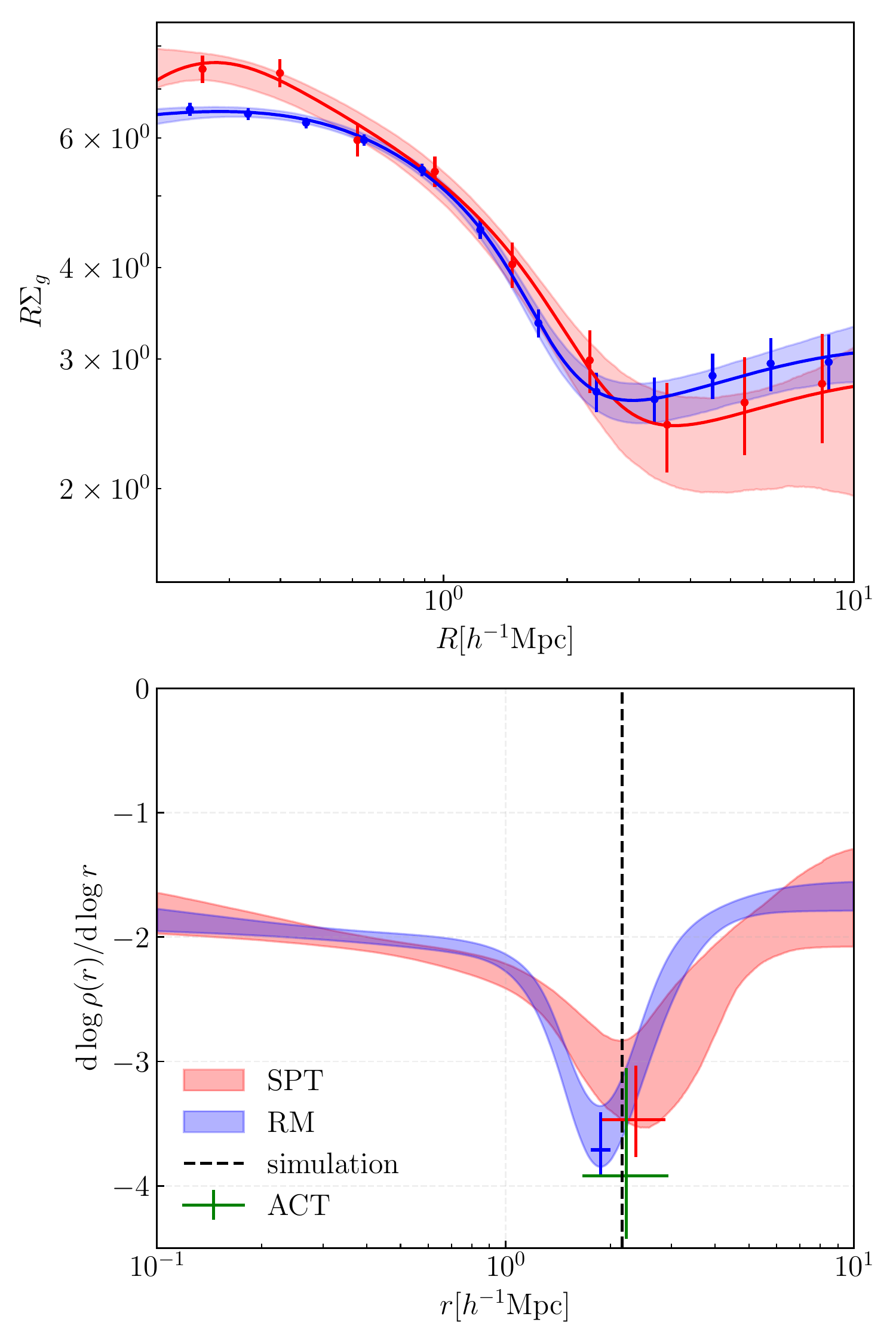}
\caption{Comparison of the measured and model-fitted galaxy profiles around SPT (red) and DES RM (blue) clusters.
In the top panel, we show the measured 2D density profiles (points with errobars), the best-fit model curve (solid line) and 1$\sigma$ range of the fitted profile (bands) of each cluster sample in the corresponding color. 
In the bottom panel, the 1$\sigma$ ranges for the fitted logarithmic slope (bands),  $r_{\rm sp}$ (horizontal errorbars) and the slope at $r_{\rm sp}$ (vertical errorbars) for each cluster sample are shown.
We also show the 1$\sigma$ ranges for $r_{\rm sp}$ and the slope at $r_{\rm sp}$ for ACT clusters with the green cross.
The black dashed line shows $r_{\rm sp}$ from the simulation. The RM clusters exhibit a smaller $r_{\rm sp}$ by $\sim$2$\sigma$ than that of the simulation, consistent with previous studies with RM clusters.
}
\label{fig:sigmag_sptdes}
\end{figure}

\subsection{Comparison with redMaPPer clusters}
\label{sec:rm-comp}

In \Fref{fig:2nd-deriv} and \Fref{fig:sigmag_sptdes} we compare the measurements around the SZ-selected (SPT and ACT) clusters to those around the mass-matched DES RM clusters described in Sec.~\ref{sec:rm-dat}. In \Fref{fig:sigmag_sptdes}, we plot $R\Sigma_{\rm g}$ to highlight differences between the profiles. Due to the larger uncertainty in the ACT measurement, we focus on a comparison between the SPT and RM clusters in this section.

The RM-selected clusters prefer a smaller splashback radius --- $r_{\rm sp} = 1.88^{+0.13}_{-0.12} h^{-1}{\rm Mpc}$ --- compared to the SPT clusters, but this difference is not statistically significant ($\sim$$1\sigma$); the $r_{\rm sp}$ of SPT clusters lies on the top of the simulation value and is $\sim$$1\sigma$ larger than that of RM. On the other hand, the splashback radius from RM is smaller than the expectation from simulations by $\sim 2\sigma$, consistent with earlier results \citep{More2016,Baxter2017,Chang2017}. Our results thus show that the difference between the predicted and observed splashback radii with RM clusters persists at the high-mass end.

While SPT and RM clusters show $r_{\rm sp}$ values that are statistically consistent, there are significant differences between the galaxy surface density profiles of the two samples. For $R \lesssim 0.5\,{\rm Mpc}$, the RM clusters exhibit a smaller galaxy surface density and a shallower profile than the SPT clusters. This may be due to differences in the miscentering distributions of the two samples (\ref{sec:model}), as the inferred 3D logarithmic slope is consistent between the RM and SPT samples.

Additionally, the minimum of the logarithmic derivative of the profile of the RM clusters is lower than 
that of the SPT-selected clusters near splashback. The profile shape  can be further quantified by the third derivatives of the surface density profiles (second derivatives of the slope profiles) at the splashback radius. The results are shown in \Fref{fig:2nd-deriv}. The RM-selected clusters prefer profiles with larger third derivatives at $r_{\rm sp}$ (1-$\sigma$ range of [18.0, 60.8]) than the SZ-selected clusters ([6.1, 31.7] for SPT and [5.3, 38.4] for ACT) and the particle profile ([9.5, 11.9]) in the simulations.
It indicates that at $r_{\rm sp}$ the slope in the logarithmic derivative of RM clusters changes much faster than those of SPT clusters and simulation particles.
This measurement is consistent with \Fref{fig:sigmag_sptdes}, which shows that the RM-selected clusters have a narrow minimum in their logarithmic derivative profiles.

\begin{figure}
	\includegraphics[width=1.0\linewidth, trim= 0in 0.0in 0in 0in,clip]{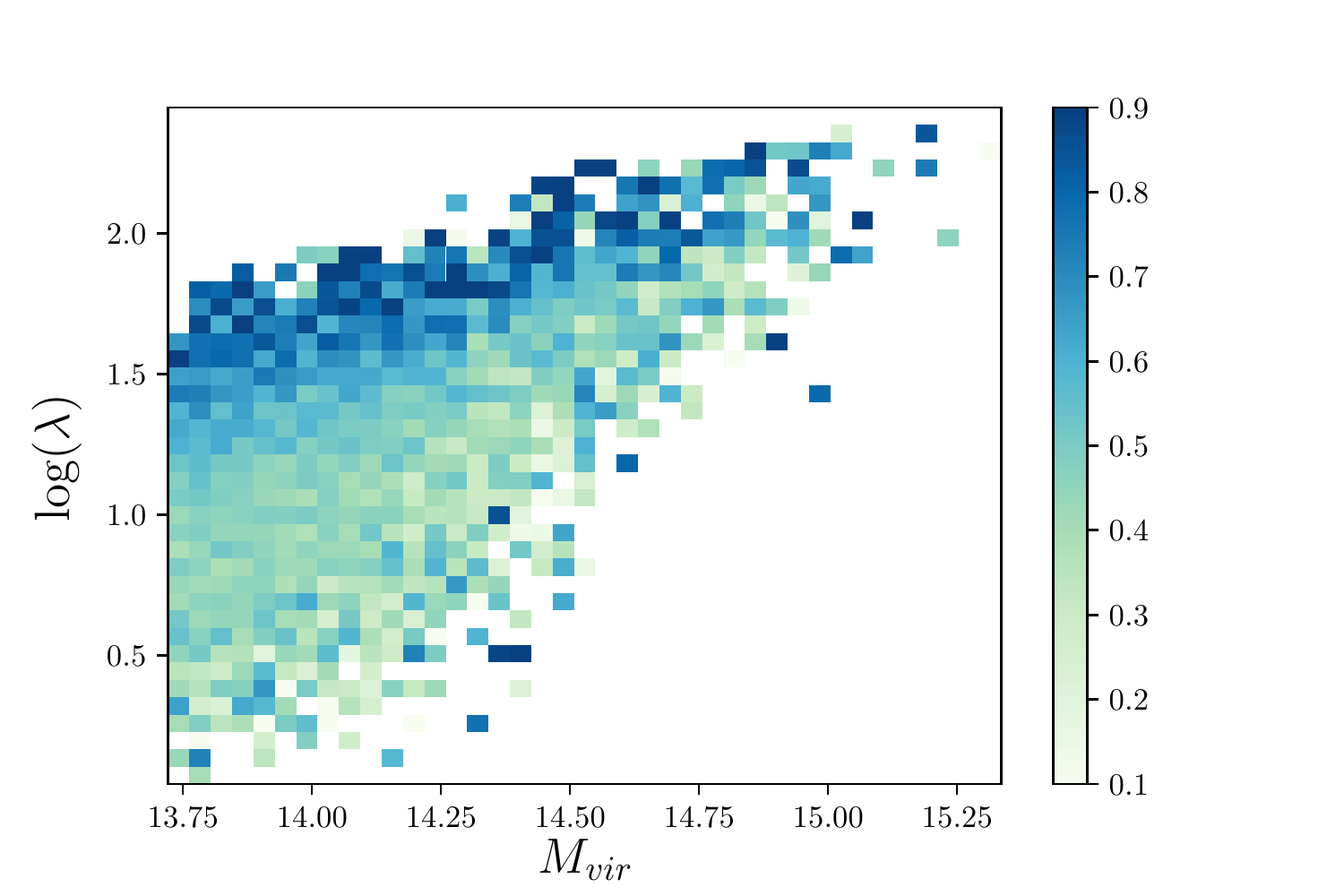}
    \caption{Mass-Richness distribution of halos from RM mock catalogs obtained from Buzzard simulations. The samples are color-coded by the cosine of the orientation angle. High richness galaxies have major axes preferentially oriented towards the observer.
}
\label{fig:mlambda}
\end{figure}

\begin{figure}
\includegraphics[width=1.0
\linewidth, trim= 0.0in 0.0in 0in 0in,clip]{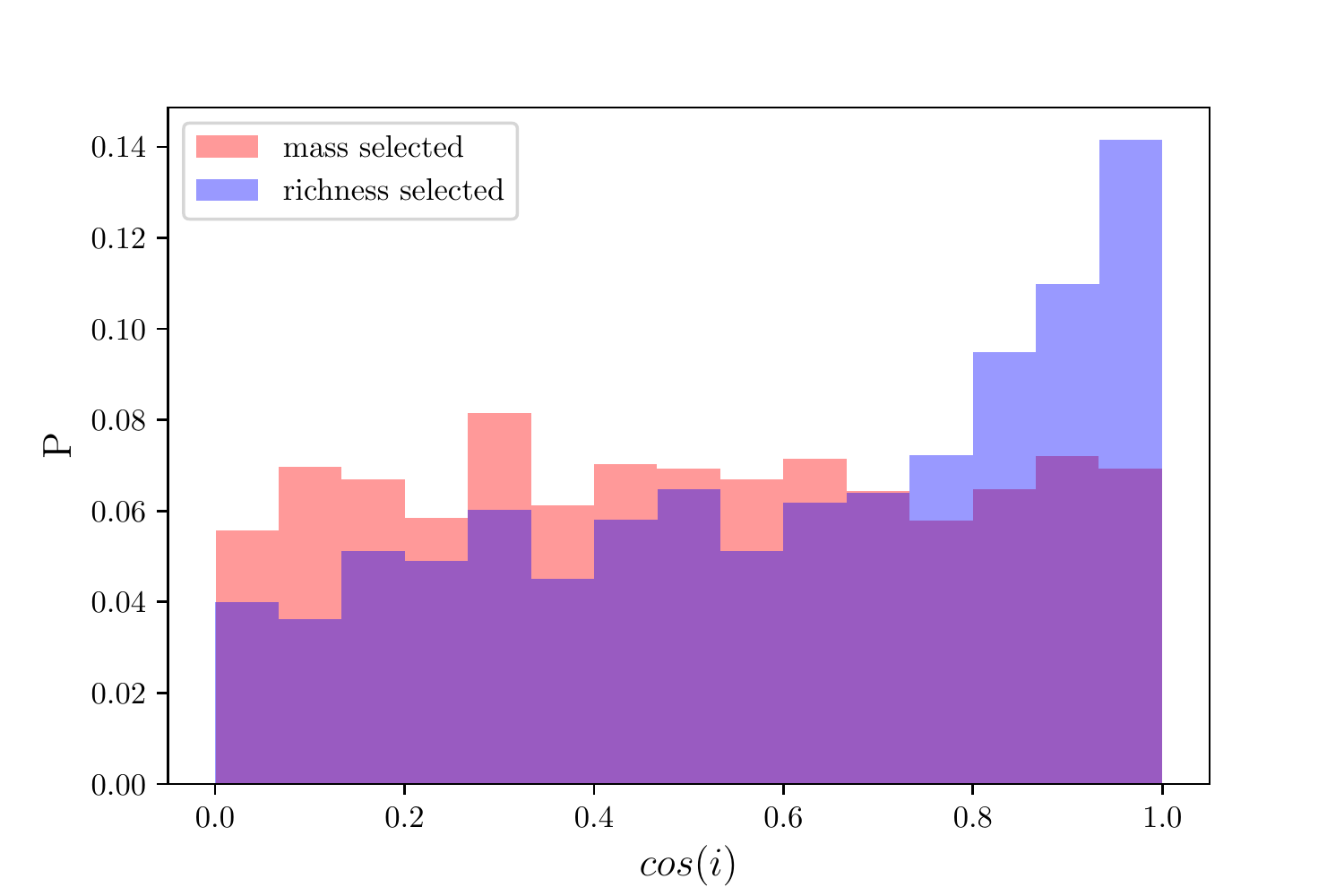}
    \includegraphics[width=1.0\linewidth, trim= 0.0in 0.0in 0.0in 0in,clip]{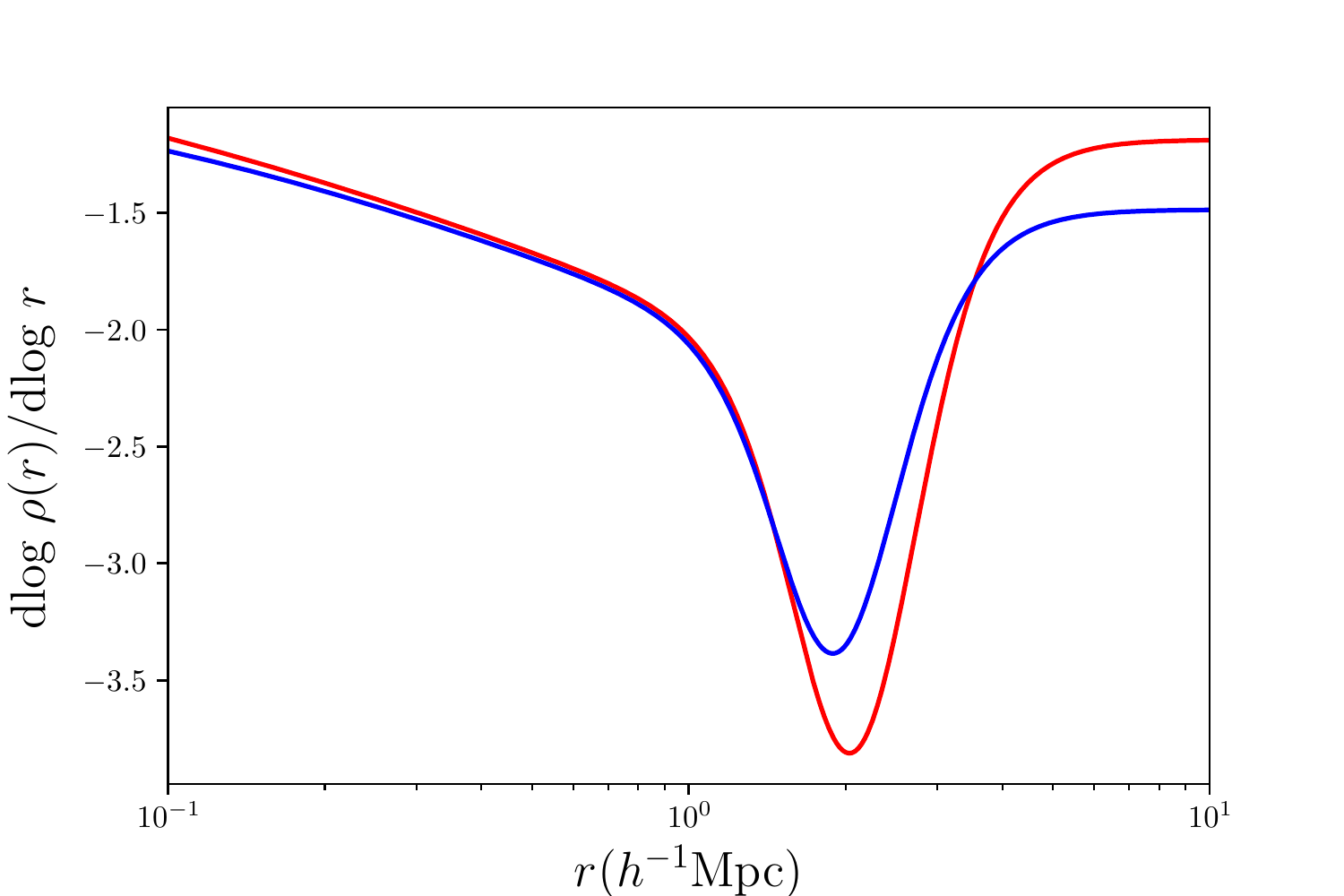}
\caption{Top: The distribution of the orientation angle for samples with the same mean mass $\langle M_{200m}\rangle = 5.09\times10^{14} M_\odot h^{-1},$ selected from Buzzard either by mass (red) or richness (blue). Bottom: Profile slopes measured in the two cases: red corresponds to the mass-selected sample and blue corresponds to the richness-selected sample. The shift in $r_{\rm sp}$ between the two cases is $\sim$6\%.   
}
\label{fig:orientation}
\end{figure}

One possible explanation for the difference in the splashback radius between the RM and simulated clusters is orientation bias introduced in RM cluster selection \citep{2014MNRAS.443.1713D}. Because halos are assigned a richness, $\lambda$, based on the overdensity of red galaxies within an aperture, any selection of a halo sample that is based on a richness cut is likely to include halos that have their major axis preferentially oriented towards the line of sight. The splashback radius can be different along different axes in a triaxial halo. Therefore we may expect that if we are systematically looking down the major axis and stacking halos based on a richness cut-off, the 2D splashback radius may shift to a smaller radius. We note that this is a different effect from what is discussed in \citet{Busch2017}, as in that study it is suggested that the random fluctuation of the galaxy distribution relative to the dark matter distribution, coupled with the richness selection, is the source of the splashback feature, whereas here the galaxy distribution is aligned with the dark matter, but the richness selection preferentially selects dark matter halos that are oriented along the line-of-sight.

To quantify the orientation bias that may be present in our sample of clusters we use RM mocks generated using the Buzzard simulations \citep{DeRose2019}. These are a set of dark matter only, $\Lambda$CDM, N-body simulations that simulate the DES lightcone, by painting galaxies on dark matter particles using ADDGALS (for details see \citet{DeRose2019} and \citet{Busha:2008uyz}). For each halo associated with a RM cluster, we use the dark matter particles within $R_{200m}$ to calculate the reduced inertia tensor \citep[see e.g.,][]{Osato2018}, and its largest eigenvalue and the corresponding eigenvector are associated with the major axis of the halo. The cosine of the angle between this major axis and the line-of-sight direction, $\cos(i)$, quantifies the orientation of the halo. For a sample of randomly selected halos, $\cos(i)$ follows a uniform distribution.

\Fref{fig:mlambda} shows the distribution of cosine of the angle that the major axis of a cluster makes with the line-of-sight direction as a function of the mass and richness of Buzzard halos. The top panel of \Fref{fig:orientation} shows the comparison of the distribution of $\cos(i)$ for samples of clusters with the same mean mass but selected by applying a lower mass threshold in one case and a richness threshold in the other.  As is evident from this figure, there is a significant orientation bias in the mock RM sample(see also \citet{zhouweninprep} for a detailed study of orientation bias). 

The bottom panel of \Fref{fig:orientation} shows the splashback radius measured for the two samples using the Buzzard halos. The richness for these halos have been obtained by running the RM algorithm on halo centers. While the median of $\cos(i)$ in the richness selected sample shifts to 0.67 the splashback radius changes only by about $6\%$. To get effects on the order of $20\%$ the median orientation angles of the two samples must differ by $~0.5$. We also test the effect of orientation bias in our fiducial simulation. 
We measure the orientation angle, $i$, in the MDPL2 simulations and then select halos to reproduce the orientation distribution of RM in Buzzard simulations. 
It shows a similar amount of shift in $r_{\rm sp}$ as in Buzzard and thus also cannot completely explain the shift in the splashback radius between RM and SZ clusters.
Nevertheless, we caution that this exercise is based on simulations in which the characteristics of the galaxies may not completely match our data.
This means that it is still possible that the \textit{quantitative} level of this selection effect is closer to what we observe than what is shown in the simulation.

Another potential explanation for the discrepancy between the RM and simulation measurements is a bias in the cluster mass-observable relations.  If, for instance, the mass-richness relationship from \citet{mcclintock2019} were biased to high masses, then the true mass of the RM sample would be lower than we have inferred.  Naively, this is not an impossible explanation, since the number density of our RM sample is higher than that of our SPT sample, which could be consistent with a mass bias.  However, note that to explain the observed 10\% discrepancy in the splashback radius relative to simulations would require a roughly 30\% bias in the mass-richness relation, significantly larger than the 5\% uncertainty reported by \citet{mcclintock2019}.

We can further test bias in the mass-richness relation as a possible explanation for the observed splashback discrepancy by abundance matching the RM sample to the SZ sample.  This approach has the advantage of being independent of the mass-observable relations of the two samples.  We select RM clusters with $\lambda > \lambda_{\rm am}$, choosing $\lambda_{\rm am}$ such that the number density of RM clusters matches that of SPT clusters.  For the abundance matched RM sample, we find that the mean mass is $\bar{M}_{\rm 200m} = 6.27\times 10^{14} h^{-1} M_{\odot}$, 18\% larger than the mean mass of the fiducial RM sample.  Fitting the galaxy density measurement around the abundance matched sample, we find that the inferred splashback radius increases to $2.03^{+0.15}_{-0.17}$, 8\% higher than the fiducial RM measurement (see \Fref{fig:sigmag_sptdes}.), and $0.7\sigma$ ($0.3\sigma$) below the prediction in simulations with subhalos (particles).  While the $r_{\rm sp}$ tension relative to simulations is therefore reduced, the mass-richness relationship from \citet{mcclintock2019} would need to be in error by roughly $3.6\sigma$ for this to be the full explanation of the RM-simulation discrepancy.  The distribution of third derivatives at splashback for the abundance matched sample is shown in Fig.~\ref{fig:2nd-deriv}.  We find that the abundance matched sample also has a significantly narrower splashback feature than the simulated clusters.  Consequently, even if a large bias in the mass-richness relationship were the explanation for the low $r_{\rm sp}$ measured for RM clusters, the shape of the splashback feature for these clusters would still be discrepant with simulations.

In summary, we do not have evidence that any single factor explains the amplitude of the discrepancy between the measured $r_{\rm sp}$ in the simulated and RM clusters.
However, tests with simulations point to plausible consequences of RM selection that contribute to the difference. The simplicity of the SZ-selection function, on the other hand, makes it a more robust (albeit lower SNR so far) measurement. This type of analysis will be much more powerful with the larger SZ-selected cluster samples expected from ongoing and future surveys \citep{eRosita,SPTPol,SPT3G,advACT,S4CMB2016,SO2018}.

\begin{figure*}
\centering
\includegraphics[width=0.9\linewidth]{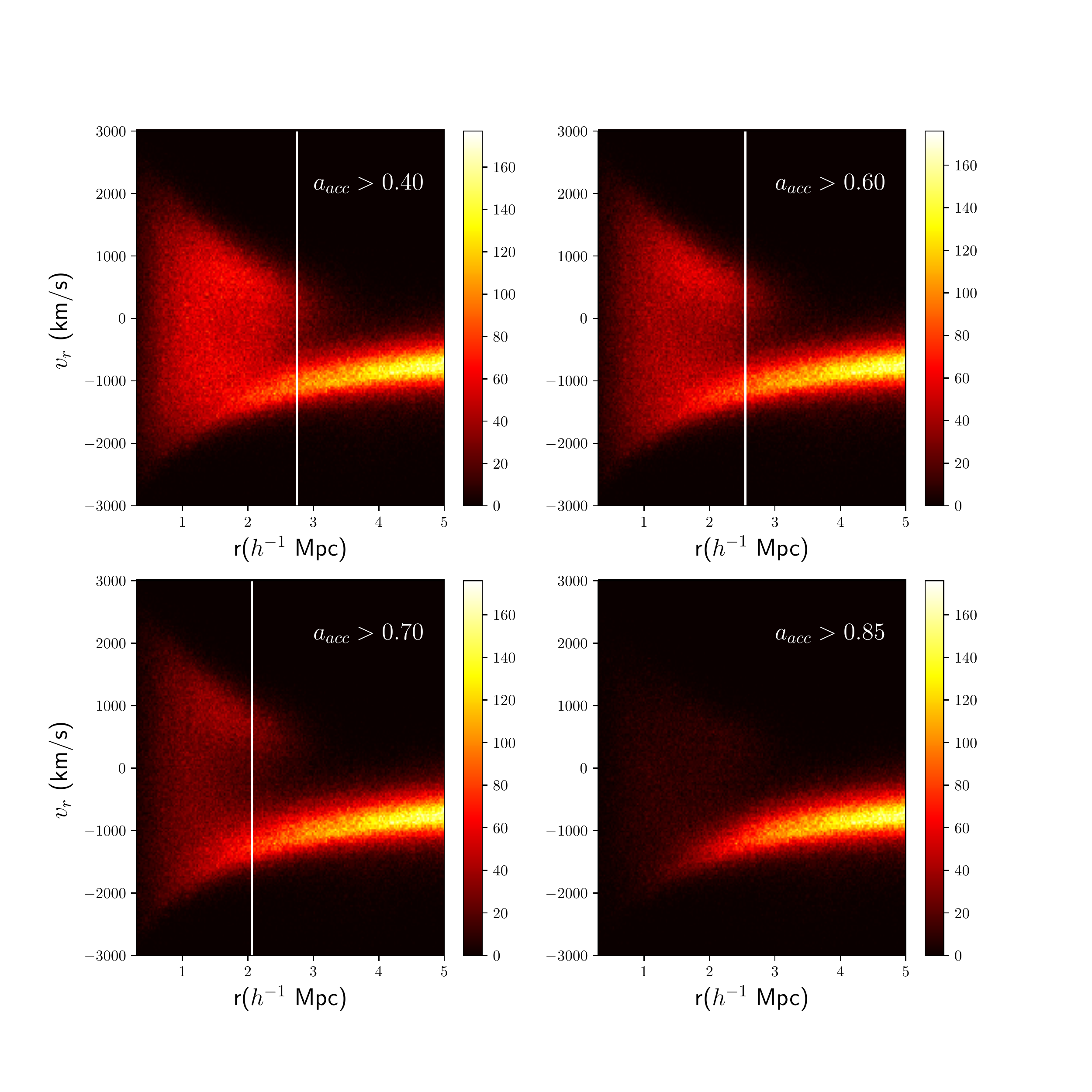}
\caption{Subhalos accreted by a cluster halo at different times, plotted in phase space. The four panels show the phase space distribution of all subhalos accreted on to hosts with mean mass $M_{\rm 200m}=6.17\times 10^{14}M_{\odot} h^{-1}$ at times later than an ``accretion time'', when the subhalo crossed into 3.8$h^{-1}$ Mpc. Each panel gives the scale factor $a_{acc}$ corresponding to the accretion time. These are from the zoom-in simulations Rhapsody \citep{Rhapsody_heidi}. The white vertical lines indicate the minimum of the slope of the 3D density profile in each of the four cases except in the lower-right panel in which there is no distinct splashback feature. The lower panels show that subhalos that are accreted late, suggestive of blue galaxies, have not had time to splash back. The minimum in their slopes in 3D density profiles, if it exists, is not a true splashback.
}
\label{fig:phase}
\end{figure*}

\begin{figure}
\centering
\includegraphics[width=0.99\linewidth]{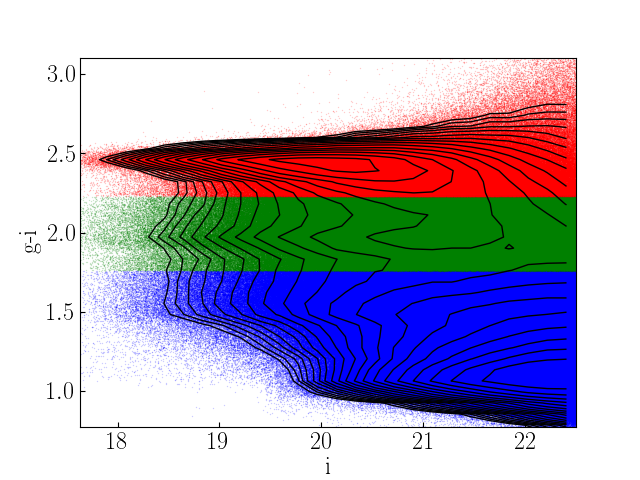}
\includegraphics[width=0.99\linewidth]{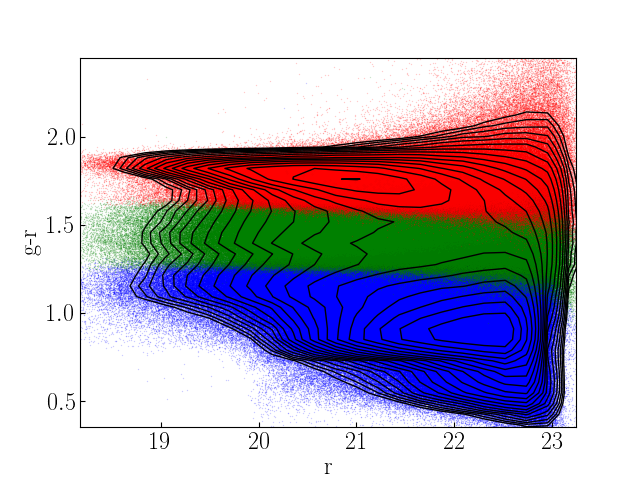}
\includegraphics[width=0.99\linewidth]{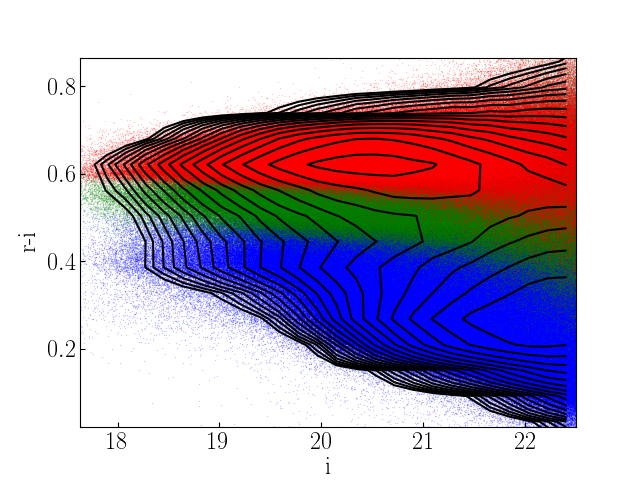}
\caption{\textit{top:} Galaxy color distribution in \textit{g-i} color-magnitude space with the galaxy density contour over-plotted, in the redshift bin $z$=[0.475,0.500] which encompasses the mean redshift of our fiducial SPT cluster sample.
The ``red sequence'' and the ``blue cloud'' are clearly seen, as well as the ``green valley''. Accordingly, as described in Sec.~\ref{sec:color_def}, we define the red, green and blue galaxies so that they consist of 20\%, 20\% and 60\% of the entire galaxy population, respectively.
\textit{middle, bottom:} The corresponding red/green/blue galaxy distributions over \textit{g-r} (middle) and \textit{r-i} (bottom) color-magnitude space. One can see the separation in \textit{g-i} color space results in a reasonable color selection also in other color spaces.
}
\label{fig:color}
\end{figure}

\begin{figure}
\centering
\includegraphics[width=1.0\linewidth]{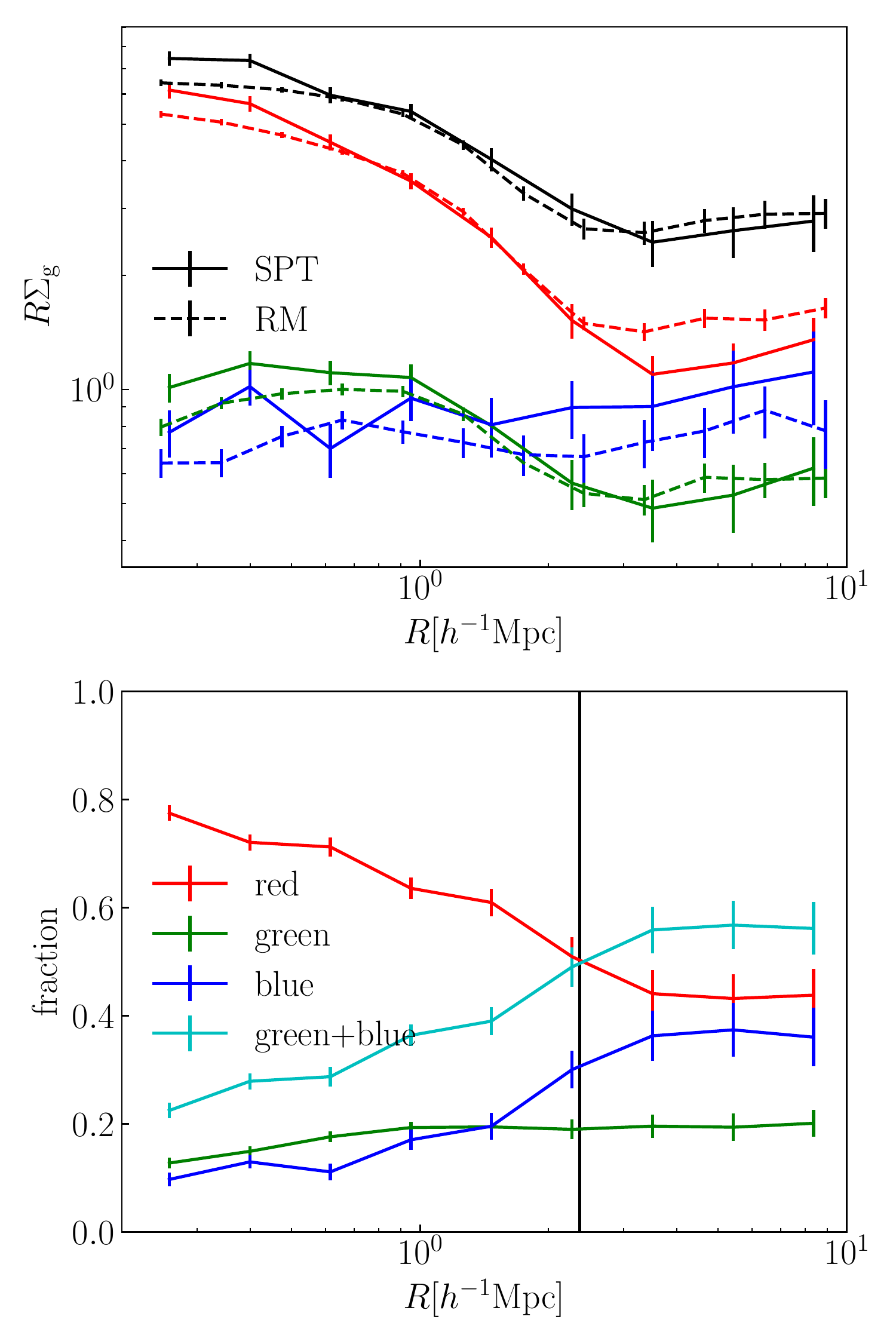}
\caption{Red, green and blue galaxy density profiles around the SPT and RM clusters (top) and the fraction of the corresponding color galaxies for SPT clusters with respect to all galaxies in that radial bin (bottom). The vertical line shows the location of the 3D splashback radius for SPT clusters. 
}
\label{fig:clr_frac}
\end{figure}

\begin{figure}
\centering
\includegraphics[width=1.0\linewidth]{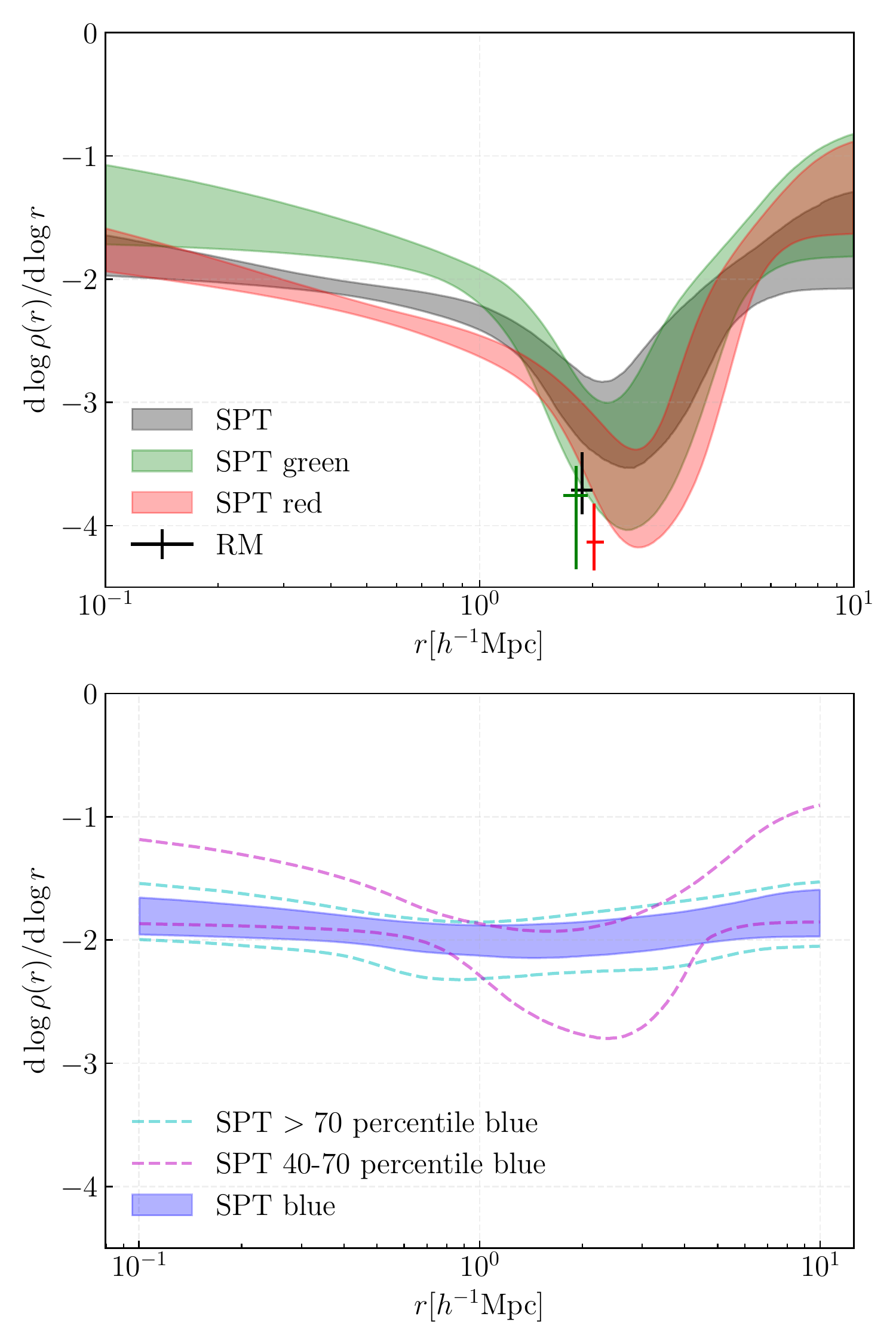}
\caption{
\textit{Top:} The 68\% confidence ranges of the 3D logarithmic derivative of the fitted galaxy density profiles with all galaxies (black), red galaxies (red) and green galaxies (green) around SPT clusters. Also shown in black, red and green crosses are the 68\% ranges for $r_{\rm sp}$ (horizontal errorbar) and the logarithmic slope at $r_{\rm sp}$ (vertical errorbar) around RM clusters.
\textit{Bottom:} similar but for blue galaxies (blue), and the redder half (40-70 percentile in color, dashed magenta) and the bluer half (>70 percentile in color, dashed cyan) of the blue galaxies.
}
\label{fig:prof_color}
\end{figure}

\subsection{Quantitative comparison of profiles}
\label{sec:comparison_profile}

In addition to comparing the inferred splashback radii of the different cluster samples, we can also directly compare the profile measurements for these clusters. Below, we perform comparisons between the measured and simulated profiles, and between the SZ-selected and RM cluster profiles. Again, due to the larger uncertainty in the ACT measurement relative to that of SPT, we focus on SPT clusters for the SZ-selected sample.

We use a $\chi^2$ test to evaluate consistency between the various profile measurements.  Since the uncertainty on the simulated profiles is small compared to the uncertainty in the measured profiles, we ignore this source of uncertainty in comparisons to the simulated profiles.  However, when comparing the measured and simulated profiles, we must account for differing normalizations of these profiles. We therefore define a $\chi^2$ via
\begin{equation}
\chi^2(\alpha) = \left( \vec{d} - \alpha \vec{s} \right)^T \mathbf{C}^{-1} \left( \vec{d} - \alpha \vec{s} \right),
\end{equation}
where $\vec{d}$ represents the cluster profile measured in data, $\vec{s}$ represents the measurements in simulations, $\alpha$ is a free parameter, and $\mathbf{C}$ is the covariance matrix of the measurements. We expect the minimum $\chi^2$ obtained by varying $\alpha$ to be $\chi^2$ distributed with degrees of freedom equal to $\nu = N_{R} - 1$, where $N_{R}$ is the number of radial bins. We quantify tension between the profile measurements by reporting the probability to exceed (p.t.e.) the measured minimum $\chi^2$. Since the galaxy profile is not expected to trace the particle or subhalo profiles at very small scales, we restrict the analysis in this section to scales $R > 0.5\,h^{-1}{\rm Mpc}$.  

In the comparison of galaxy profiles around SPT clusters to the simulated particle profile we find ${\rm p.t.e.} = 0.004$, indicating that the particle profile in the CDM simulation is significantly different from the measured galaxy profile. When comparing to the subhalo profile, we find ${\rm p.t.e.} = 0.2$, indicating that the galaxy profile measurements around SPT clusters are consistent with the simulated subhalo profiles.

We repeat the $\chi^2$ test described above for the RM-selected clusters.  Comparing to the particle profile, we find ${\rm p.t.e.} = 2\times 10^{-4}$, indicating that the particle profile in simulations is not consistent with the measured galaxy profile. When using the subhalo profile (and again restricting to scales above $R > 0.5\,h^{-1}{\rm Mpc}$), we find ${\rm p.t.e.} = 0.03$. This low p.t.e. value indicates mild tension between the profiles of the RM clusters and the subhalo profile in simulations, unlike the case of SPT clusters.

Finally, we can also compare the SZ-selected clusters directly to the RM-selected clusters. In this case, allowing for a free normalization parameter is unnecessary, and we can compute $\chi^2$ via $\chi^2 = (d_{\rm SZ} - d_{\rm RM})^T \mathbf{C}^{-1} (d_{\rm SZ} - d_{\rm RM})$, with $\nu = N_R$. In this comparison, we find ${\rm p.t.e} = 0.6$ when including only scales $R > 0.5\,h^{-1}{\rm Mpc}$, indicating that the SPT and RM cluster profiles are statistically consistent over these scales. When including all measured scales, however, we find ${\rm p.t.e.} = 0.01$, indicating tension. In agreement with \Fref{fig:sigmag_sptdes}, the tension between the RM and SPT cluster profiles is driven by the smallest radial bins, which may be related to different levels of miscentering for the two samples.

\subsection{Splashback as a function of galaxy color}
\label{sec:comp-color}

Clusters are associated with a high density of red and elliptical galaxies that have very little to no star formation \citep{Dressler:1980wq,Dressler:1984kh,Balogh:2000sf,Poggianti:1999xh,Oemler:1974yw}. The quenching of star formation within a cluster may be related to intra-cluster processes like ram-pressure stripping \citep{Abadi:1999qy,Gunn1972}, strangulation \citep{Larson:1980mv}, or galaxies may quench due to intrinsic processes related to their age \citep[see e.g.][]{Wetzel2013, vonderLinden2010, Brodwin2013, Ehlert2014, Wagner2015}. In either case the fraction of red and blue galaxies should show a sharp variation at the splashback radius as it is the physical boundary that separates the region of space with galaxies on orbits (which are older and also more likely to be quenched) from those on pure infall at larger radii. This sharp transition near the edges would not be present if the one-halo term was NFW-like and continued onto the two-halo regime smoothly, and is therefore evidence for an abrupt cut-off of the virialized region at splashback \citep{Baxter2017}. 

The purpose of our analysis here is two-fold. First, we seek to improve on the red/blue selection of galaxies in \citet{Baxter2017} and see whether this improvement changes the conclusion in that paper. Second, we seek to compare the color-split profiles for SZ-selected clusters with that of RM-selected clusters, since the latter cluster finder specifically uses a subset of the red galaxies to find the cluster and may be susceptible to biases that depend on galaxy colors.

Furthermore, if galaxies stop forming stars during their orbits within a cluster, i.e. if quenching is mainly due to intra-cluster astrophysical processes, then the longer a galaxy has been inside a cluster the more likely it is to be red. So the color of a galaxy should be indicative of how long it has been inside the cluster.
Imagine a sample of blue galaxies falling into the cluster: if quenching begins on entry into the cluster and no blue galaxies survive until pericentric passage, then the density profile of the blue galaxies should not show any splashback, since none of them are able to reach the first apocenter of their orbit to form a splashback shell. 
Furthermore, since they are still on their first infall passage, the density profile of those blue galaxies would be consistent with a pure power law. Fig.~\ref{fig:phase} demonstrates how the slope of the 3D density profile traces discontinuities in phase space. The four panels show the phase space distribution of subhalos that have been inside the cluster for different amounts of time. Galaxies in the infall stream do not show a splashback-like minimum, those that have completed at least one crossing show a minimum at splashback, while those that have not reached splashback but have crossed pericenter show a slope minimum at a location smaller than the splashback radius. In this paper, we study the logarithmic slope and surface density for galaxies of different colors around massive SZ clusters, which contains information about their accretion histories.

\subsubsection{Defining galaxy colors}
\label{sec:color_def}
To make galaxy color selections, we divide galaxies in the redshift range [0.25,0.7] into bins with width $\Delta z=0.025$. Then in each redshift bin, we divide galaxies into three percentile ranges in \textit{g-i} color. In \Fref{fig:color}, we show the result in the redshift bin [0.475,0.500], which includes the mean redshift of the SPT cluster sample. The galaxy density contours are over-plotted in color-magnitude space. From the density contours, one can observe a bimodal distribution consisting of the red sequence and the blue cloud, as well as the green valley between them \citep{baldry2004,schawinski2014}. Furthermore, the red sequence approximately includes the reddest 20\% of the galaxies, the green valley the next 20\% and the blue cloud the next 60\%. We adopt the aforementioned threshold (20\%/20\%/60\% division) as our color definition, and call them the `red', the `green' and the `blue' galaxies, respectively. Note that the fraction of blue galaxies drops significantly inside clusters (\Fref{fig:clr_frac}), as we discuss further below.

The red/green/blue fractions also evolve moderately with redshift. For example, in the lowest redshift bin of $z=[0.250,0.275]$, the red sequence covers ${\sim}10-15\%$ of galaxies, whereas the blue cloud includes ${\sim}65-70\%$ of them. However we have adopted the 20\%-20\%-60\% separation in \textit{g-i} color over the entire redshift range.
We have checked that our color split also results in a reasonable separation of galaxies in other color spaces (see the lower two panels of \Fref{fig:color}).

\subsubsection{Galaxy density profiles in different color bins}
The measured galaxy surface density profiles and their corresponding logarithmic derivatives, in different color bins, are shown in \Fref{fig:clr_frac} and \Fref{fig:prof_color}, respectively.
In the upper panel of \Fref{fig:clr_frac}, we plot the surface density profiles of all/red/green/blue galaxies around SPT (solid lines) and RM clusters (dashed lines). 
In the bottom panel, we calculate the fraction of the red, green and blue galaxies as in \citet{Baxter2017} by dividing each profile by the profile with all galaxies. 
It shows a sharp upturn (downturn) of the red (blue) fraction around the splashback radius, similar to the results in \citet{Baxter2017}. Note that the upturn of the red fraction starts at a higher radius ($\sim$3 $h^{-1}$Mpc) than $r_{\rm sp}$. This may be attributed to the width of the splashback region, as not all galaxies turnaround exactly at the location of the minimum \citep{2017ApJS..231....5D,mansfield2017}.
In addition, some galaxies may start quenching before they infall onto the cluster by pre-processing in the infalling galaxy groups \citep[e.g.,][]{zabludoff96,behroozi14,bianconi18}.

In the top panel of \Fref{fig:prof_color}, we plot 68\% confidence ranges of the 3D logarithmic derivative of the fitted galaxy profiles around the SPT clusters, with all galaxies (black), green galaxies (green) and red galaxies (red), respectively. 
Red galaxies in SPT clusters display splashback features that are slightly deeper, consistent with \citet{Baxter2017}. 
In the same figure, we also show 68\% ranges of the $r_{\rm sp}$ (horizontal errorbars) and the corresponding logarithmic slope (vertical errorbars) around the RM clusters, with the same color bins. The RM clusters exhibit a similar trend of the splashback feature across different galaxy colors.

In the bottom panel of \Fref{fig:prof_color}, the 3D logarithmic derivative profiles of blue galaxies around the SPT clusters are plotted in blue.
Although the data is somewhat noisy, one can see that the blue galaxy profiles are consistent with a pure power law not exhibiting any evidence of splashback feature, which may indicate the blue galaxies are still in their first orbital passage inside the clusters.
We further split blue galaxies into halves: the redder half (40th-70th percentile, dashed magenta lines) and the bluer half (>70th percentile, dashed cyan lines).

These results suggest that infalling galaxies do not remain blue beyond their first pericentric passage, though perhaps not all star formation is quenched by the first apocentric passage as green galaxies do show a splashback feature close to the red galaxies. However, we defer a more detailed study of the profiles of galaxies split on color and galaxy type to when larger SZ-selected samples are available. It would also be very interesting to have cluster/galaxy samples that extend to $z>1$ where rapid quenching is expected \citep{Brodwin2013,Ehlert2014,Wagner2015}, and to use measures of star-formation beyond galaxy color to constrain quenching timescales. 

\section{Discussion}
\label{sec:discussion}

We have presented measurements of galaxy profiles around SZ-selected galaxy clusters and used these measurements to characterize the splashback feature around these clusters. The SZ-selected sample has the advantage that it is likely to be closer to a mass-selected sample than optically selected cluster samples. Also, it provides an independent check on previous measurements of the splashback feature, which are all based on optically selected cluster samples. We used publicly available cluster samples from the SPT-SZ survey and the ACTPol survey for this study. 
These samples include 256 and 89 clusters in the DES footprint, respectively. 
The clusters were cross-correlated with galaxies from the DES Year 3 dataset.
We detect the splashback feature, inferred from a sharp decline in the galaxy density profile, with high significance for both cluster samples. 
The detection of the splashback feature is confirmed by the slope of the collapsed inner profile ($\rho_{\rm coll}$) being much steeper than that from an NFW profile ($-3$), by $\sim$$2\sigma$. 
When comparing to the MDPL2 N-body simulations, both the location and the amplitude (the steepest slope) of the splashback features in the two cluster samples agree with the simulations at the 1$\sigma$ level.

To connect with previous studies based on optically selected clusters, we match the mean mass and redshift distribution of clusters in the DES Y3 RM (redMaPPer) cluster sample to the SPT and ACT cluster samples and measure the location of the splashback feature in this mass-matched RM sample.
We find that the location of the splashback feature is at a smaller radius than simulations, consistent with previous studies with RM clusters \citep{More2016,Baxter2017,Chang2017}. 
The size of this tension is too large to be easily explained by bias in the mass-richness relation. This suggests that RM clusters are likely affected by systematic effects that push the splashback radius to smaller values. We also investigate the possibility that projection effects coupled with the triaxiality of the clusters contribute to the selection effect in the RM clusters. From simple simulation tests, we find that these effects do contribute to the smaller apparent $r_{\rm sp}$, but may not fully explain the level of discrepancy in $r_{\rm sp}$ between RM clusters and simulations.
Improved mock catalogs for RM clusters may provide further insights. 

\begin{table*}
	\centering
	\begin{tabular}{cccccc}
		Reference & Measurement & Sample & Mean mass[$10^{14}h^{-1}M_{\odot}$] & Mean redshift & $r_{\rm sp}/r_{\rm 200m}$ \\ \hline
%        \citet{diemerlens2017} & weak lensing & CLASH X-ray & $M_{\rm 200m}=13$ & 0.34 & >0.89 \\ 
		\citet{More2016,Baxter2017}$^*$ & galaxy profile & SDSS RM & $M_{\rm 200m}=1.9$ & 0.24 & $0.85\pm0.06$   \\
        \citet{Chang2017} & galaxy profile & DES RM & $M_{\rm 200m}=1.8$ & 0.41 & $0.82\pm0.05$ \\ 
        \citet{Chang2017} & weak lensing &  DES RM & $M_{\rm 200m}=1.8$ & 0.41 & $0.97\pm0.15$ \\ 
        This work	 & galaxy profile & DES RM & $M_{\rm 200m} = 5.3$ & 0.46 & $0.97^{+0.07}_{-0.06}$\\
		This work & galaxy profile & SPT SZ & $M_{\rm 200m} = 5.3$ & 0.49 & $1.22^{+0.26}_{-0.25}$ \\
        This work & galaxy profile & ACT SZ & $M_{\rm 200m} = 5.8$ & 0.49 & $1.11^{+0.36}_{-0.28}$ \\
        This work & particle profile & MDPL2 N-body sims & $M_{\rm 200m} = 5.3 $ & 0.49 & $1.07\pm0.02$\\
        %\citet{contigiani2018} & weak lensing & CCCP X-ray & $M_{\rm 200m}=14$ & 0.28 & $1.34^{+0.45}_{-0.26}$ 
	\end{tabular}
    \caption{The splashback radius $r_{\rm sp}$ based on the galaxy profile and lensing profile,  from previous studies as well as this paper. Note that we normalize $r_{\rm sp}$ by $r_{\rm 200m}$ for easier comparison. We note that \citet{diemerlens2017} also reported a lower limit of $r_{\rm sp}/r_{\rm 200m} > 0.89$ based on a weak lensing measurement of the CLASH X-ray cluster sample. We also note that \citet{contigiani2018} report the constraints  $r_{\rm sp}/r_{\rm 200m} = 1.34^{+0.45}_{-0.26}$ based on weak lensing measurements around X-ray selected clusters; however, they do not report a significant detection of splashback-like steepening in the cluster density profile.*The values quoted in the first row are from \citet{Baxter2017} as \citet{More2016} only reported results for cluster samples split on their $R_{\rm mem}$ parameter.  }
    \label{tab:rsp}
\end{table*}

We summarize our measurements of $r_{\rm sp}$ as well as those of previous studies in Table~\ref{tab:rsp}. The table makes clear that measurements based on galaxy density profiles around RM clusters consistently find lower splashback radii than found in simulations and in our measurements of SZ-selected clusters. The redMaPPer measurements reported here are in turn consistent with the measurements in earlier papers cited in the table which had a lower mean mass. So the comparison of RM vs. SZ clusters with theory does not appear to be related to cluster mass (or redshift). We caution that the uncertainty on the SZ measurements precludes a more definitive statement. 

Although we have focused on the location of the splashback radius in this paper, we also learn about the distribution of galaxies within massive clusters as a whole. While the location of the splashback feature is a distinctive feature that is simple to interpret physically, the overall profiles of the different samples analyzed in this paper also contain a wealth of information.
For example, the galaxy profiles around the clusters do not exactly trace the particles or the subhalos in CDM simulations; the inner profiles of the optically and SZ selected clusters are much steeper than those in the simulations (\Fref{fig:sigmag_sptsim} and \Fref{fig:sigmag_act}). 
This may require further understanding of baryonic physics; for instance, the potential of the central galaxy may be strong enough to contract particle orbits to raise the central density. Comparison with lensing profiles in  future studies will provide further insights. Furthermore, while the splashback location for the RM clusters is different from that of the SZ clusters by less than 1 $\sigma$, the overall profiles from $0.2-5 h^{-1}$ Mpc differ from each other at higher significance. We also find, in agreement with \cite{Baxter2017}, that the outer profiles beyond splashback for all the samples asymptote to a slope of $-1.5$, which is consistent with infalling matter.

We also build on the approach of \citet{Baxter2017} and measure the profiles of galaxies split by color. Red galaxies exhibit a splashback feature that is slightly deeper in the logarithmic derivative profile, while the bluest galaxies are consistent with a featureless, power law profile that is expected for galaxies that are on the first infall and have not completed one pericentric passage.
This reiterates the fact that the location of the splashback feature contains dynamical information that is otherwise difficult to obtain without velocities obtained from a redshift survey. Our results are consistent with earlier work by \citet{oman2016}, \citet{2018arXiv180609672A}, and \citet{zinger2018} who used SDSS spectroscopic data and found that blue galaxies are dominated by an infalling population. Recently \cite{lotz2018} have also used hydrodynamical simulations to show that all galaxies stop forming stars before the first pericentric passage, consistent with blue galaxies being on the first infall. It will be interesting to model the color dependence of the splashback radius to constrain the quenching time for galaxies and compare with models for galaxy evolution like those in \cite{Umachine,Wetzel2013,2013MNRAS.435.1313H}.
For example, \citet{Wetzel2013} suggests that quenching is delayed and then sudden, taking several Gyrs to complete. Given that we do not see a splashback feature in the bluest galaxies, our results seem to suggest, firstly, that the bluest galaxies do not remain as star-forming after the first pericentric passage, implying that quenching processes begin at or before pericenter; however the fact that we find green galaxies (that are star forming) do show a splashback feature suggests that complete quenching of star formation takes longer than one apocentric passage within the cluster. (see \citet{vonderLinden2010} for an earlier work on star formation inside clusters with SDSS  and also \citet{Brodwin2013}, \citet{Ehlert2014} and \citet{Wagner2015} for studies with distant galaxy clusters). Further, it should be noted that the splashback radius measured from red galaxies, excluding the population of blue galaxies, may be a better indicator of the true boundary of the virialized region of the halos. 

Moreover, the number of SZ-selected clusters is expected to increase significantly with on-going and future surveys extended to higher redshift and lower mass \citep{eRosita,SPTPol,SPT3G,advACT,S4CMB2016,SO2018}. Along with weak lensing measurements of density profiles, future cluster samples may  sharpen the trends studied here, enable applications of splashback for tests of cluster physics or cosmology \citep{2018arXiv180604302A}, allow interesting comparisons of features in gas pressure profiles from SZ measurements with features measured in matter density profiles \citep{Shi:2016xeu,Hurier:2017mfb}, and allow for detailed comparisons with optically selected cluster samples that are essential for cluster cosmology. X-ray follow up of these clusters will provide an additional avenue for understanding the evolutionary history of these objects and help establish the splashback feature as a robust probe of galaxy cluster physics. 

\section*{Acknowledgements}

We thank Surhud More for  helpful discussions, especially for suggestions on the selection of color subsamples of galaxies. We also thank Neal Dalal, Benedikt Diemer, Arka Banerjee, Amanda Farah, Teppei Okamura , Masahiro Takada and Tom Abel for stimulating discussions. 

The CosmoSim database used in this paper is a service by the Leibniz-Institute for Astrophysics Potsdam (AIP).
The MultiDark database was developed in cooperation with the Spanish MultiDark Consolider Project CSD2009-00064.

We gratefully acknowledge the Gauss Centre for Supercomputing e.V. (www.gauss-centre.eu) and the Partnership for Advanced Supercomputing in Europe (PRACE, www.prace-ri.eu) for funding the MultiDark simulation project by providing computing time on the GCS Supercomputer SuperMUC at Leibniz Supercomputing Centre (LRZ, www.lrz.de).

Funding for the DES Projects has been provided by the U.S. Department of Energy, the U.S. National Science Foundation, the Ministry of Science and Education of Spain, 
the Science and Technology Facilities Council of the United Kingdom, the Higher Education Funding Council for England, the National Center for Supercomputing 
Applications at the University of Illinois at Urbana-Champaign, the Kavli Institute of Cosmological Physics at the University of Chicago, 
the Center for Cosmology and Astro-Particle Physics at the Ohio State University,
the Mitchell Institute for Fundamental Physics and Astronomy at Texas A\&M University, Financiadora de Estudos e Projetos, 
Funda{\c c}{\~a}o Carlos Chagas Filho de Amparo {\`a} Pesquisa do Estado do Rio de Janeiro, Conselho Nacional de Desenvolvimento Cient{\'i}fico e Tecnol{\'o}gico and 
the Minist{\'e}rio da Ci{\^e}ncia, Tecnologia e Inova{\c c}{\~a}o, the Deutsche Forschungsgemeinschaft and the Collaborating Institutions in the Dark Energy Survey. 

The Collaborating Institutions are Argonne National Laboratory, the University of California at Santa Cruz, the University of Cambridge, Centro de Investigaciones Energ{\'e}ticas, 
Medioambientales y Tecnol{\'o}gicas-Madrid, the University of Chicago, University College London, the DES-Brazil Consortium, the University of Edinburgh, 
the Eidgen{\"o}ssische Technische Hochschule (ETH) Z{\"u}rich, 
Fermi National Accelerator Laboratory, the University of Illinois at Urbana-Champaign, the Institut de Ci{\`e}ncies de l'Espai (IEEC/CSIC), 
the Institut de F{\'i}sica d'Altes Energies, Lawrence Berkeley National Laboratory, the Ludwig-Maximilians Universit{\"a}t M{\"u}nchen and the associated Excellence Cluster Universe, 
the University of Michigan, the National Optical Astronomy Observatory, the University of Nottingham, The Ohio State University, the University of Pennsylvania, the University of Portsmouth, 
SLAC National Accelerator Laboratory, Stanford University, the University of Sussex, Texas A\&M University, and the OzDES Membership Consortium.

Based in part on observations at Cerro Tololo Inter-American Observatory, National Optical Astronomy Observatory, which is operated by the Association of 
Universities for Research in Astronomy (AURA) under a cooperative agreement with the National Science Foundation.

The DES data management system is supported by the National Science Foundation under Grant Numbers AST-1138766 and AST-1536171.
The DES participants from Spanish institutions are partially supported by MINECO under grants AYA2015-71825, ESP2015-66861, FPA2015-68048, SEV-2016-0588, SEV-2016-0597, and MDM-2015-0509, 
some of which include ERDF funds from the European Union. IFAE is partially funded by the CERCA program of the Generalitat de Catalunya.
Research leading to these results has received funding from the European Research
Council under the European Union's Seventh Framework Program (FP7/2007-2013) including ERC grant agreements 240672, 291329, and 306478.
We acknowledge support from the Australian Research Council Centre of Excellence for All-sky Astrophysics (CAASTRO), through project number CE110001020, and the Brazilian Instituto Nacional de Ci\^encia
e Tecnologia (INCT) e-Universe (CNPq grant 465376/2014-2).

This manuscript has been authored by Fermi Research Alliance, LLC under Contract No. DE-AC02-07CH11359 with the U.S. Department of Energy, Office of Science, Office of High Energy Physics. The United States Government retains and the publisher, by accepting the article for publication, acknowledges that the United States Government retains a non-exclusive, paid-up, irrevocable, world-wide license to publish or reproduce the published form of this manuscript, or allow others to do so, for United States Government purposes.

The South Pole Telescope program is supported by the National Science Foundation through grant PLR-1248097. Partial support is also provided by the NSF Physics Frontier Center grant PHY-0114422 to the Kavli Institute of Cosmological Physics at the University of Chicago, the Kavli Foundation, and the Gordon and Betty Moore Foundation through Grant GBMF\#947 to the University of Chicago. The McGill authors acknowledge funding from the Natural Sciences and Engineering Research Council of Canada, Canadian Institute for Advanced Research, and Canada Research Chairs program.  

The ACT project is supported by the U.S. National Science Foundation through awards AST-1440226, AST-0965625 and AST-0408698, as well as awards PHY-1214379 and PHY-0855887. Funding was also provided by Princeton University, the University of Pennsylvania, and a Canada Foundation for Innovation (CFI) award to UBC. ACT operates in the Parque Astron\'{o}mico Atacama in northern Chile under the auspices of the Comisi\'{o}n Nacional de Investigaci\'{o}n Cient\'{i}fica y Tecnol\'{o}gica de Chile (CONICYT). Computations were performed on the GPC supercomputer at the SciNet HPC Consortium and on the hippo cluster at the University of KwaZulu-Natal. SciNet is funded by the CFI under the auspices of Compute Canada, the Government of Ontario, the Ontario Research Fund - Research Excellence; and the University of Toronto. The development of multichroic detectors and lenses was supported by NASA grants NNX13AE56G and NNX14AB58G.

Work at Argonne National Laboratory was supported under U.S. Department of Energy contract DEAC02-06CH11357.

\label{lastpage}

\bibliographystyle{mn2e_adsurl}

\section*{Affiliations}
$^{1}$ Department of Physics and Astronomy, University of Pennsylvania, Philadelphia, PA 19104, USA\\
$^{2}$ Kavli Institute for Particle Astrophysics \& Cosmology, P. O. Box 2450, Stanford University, Stanford, CA 94305, USA\\
$^{3}$ SLAC National Accelerator Laboratory, Menlo Park, CA 94025, USA\\
$^{4}$ Kavli Institute for Cosmological Physics, University of Chicago, Chicago, IL 60637, USA\\
$^{5}$ Department of Astronomy and Astrophysics, University of Chicago, Chicago, IL 60637, USA\\
$^{6}$ Department of Astronomy, Cornell University, Ithaca, NY 14853, USA\\
$^{7}$ Argonne National Laboratory, 9700 South Cass Avenue, Lemont, IL 60439, USA\\
$^{8}$ Universit\"ats-Sternwarte, Fakult\"at f\"ur Physik, Ludwig-Maximilians Universit\"at M\"unchen, Scheinerstr. 1, 81679 M\"unchen, Germany\\
$^{9}$ Department of Physics, Stanford University, 382 Via Pueblo Mall, Stanford, CA 94305, USA\\
$^{10}$ Astrophysics \& Cosmology Research Unit, School of Mathematics, Statistics \& Computer Science, University of KwaZulu-Natal, Westville Campus, Durban 4041, South Africa\\
$^{11}$ Enrico Fermi Institute, The University of Chicago, Chicago, IL 60637, USA\\
$^{12}$ Brookhaven National Laboratory, Bldg 510, Upton, NY 11973, USA\\
$^{13}$ Department of Physics, University of Arizona, Tucson, AZ 85721, USA\\
$^{14}$ Max Planck Institute for Extraterrestrial Physics, Giessenbachstrasse, 85748 Garching, Germany\\
$^{15}$ Department of Physics, The Ohio State University, Columbus, OH 43210, USA\\
$^{16}$ Joseph Henry Laboratories of Physics, Jadwin Hall, Princeton University, Princeton, NJ 08544, USA\\
$^{17}$ Fermi National Accelerator Laboratory, P. O. Box 500, Batavia, IL 60510, USA\\
$^{18}$ LSST, 933 North Cherry Avenue, Tucson, AZ 85721, USA\\
$^{19}$ CNRS, UMR 7095, Institut d'Astrophysique de Paris, F-75014, Paris, France\\
$^{20}$ Sorbonne Universit\'es, UPMC Univ Paris 06, UMR 7095, Institut d'Astrophysique de Paris, F-75014, Paris, France\\
$^{21}$ Canadian Institute for Theoretical Astrophysics, University of Toronto, 60 St. George St., Toronto, ON M5S 3H8, Canada\\
$^{22}$ Department of Physics and Astronomy, University of Missouri, Kansas City, MO 64110, USA\\
$^{23}$ Department of Physics \& Astronomy, University College London, Gower Street, London, WC1E 6BT, UK\\
$^{24}$ Department of Physics, University of Chicago, Chicago, IL 60637, USA\\
$^{25}$ Centro de Investigaciones Energ\'eticas, Medioambientales y Tecnol\'ogicas (CIEMAT), Madrid, Spain\\
$^{26}$ Laborat\'orio Interinstitucional de e-Astronomia - LIneA, Rua Gal. Jos\'e Cristino 77, Rio de Janeiro, RJ - 20921-400, Brazil\\
$^{27}$ Department of Astronomy, University of Illinois at Urbana-Champaign, 1002 W. Green Street, Urbana, IL 61801, USA\\
$^{28}$ National Center for Supercomputing Applications, 1205 West Clark St., Urbana, IL 61801, USA\\
$^{29}$ Institut de F\'{\i}sica d'Altes Energies (IFAE), The Barcelona Institute of Science and Technology, Campus UAB, 08193 Bellaterra (Barcelona) Spain\\
$^{30}$ Institut d'Estudis Espacials de Catalunya (IEEC), 08034 Barcelona, Spain\\
$^{31}$ Institute of Space Sciences (ICE, CSIC),  Campus UAB, Carrer de Can Magrans, s/n,  08193 Barcelona, Spain\\
$^{32}$ Observat\'orio Nacional, Rua Gal. Jos\'e Cristino 77, Rio de Janeiro, RJ - 20921-400, Brazil\\
$^{33}$ Department of Physics, IIT Hyderabad, Kandi, Telangana 502285, India\\
$^{34}$ Faculty of Physics, Ludwig-Maximilians-Universit\"at, Scheinerstr. 1, 81679 Munich, Germany\\
$^{35}$ Excellence Cluster Universe, Boltzmannstr.\ 2, 85748 Garching, Germany\\
$^{36}$ Department of Astrophysical Sciences, Princeton University, Princeton, NJ 08544, USA\\
$^{37}$ Department of Astronomy/Steward Observatory, 933 North Cherry Avenue, Tucson, AZ 85721-0065, USA\\
$^{38}$ Jet Propulsion Laboratory, California Institute of Technology, 4800 Oak Grove Dr., Pasadena, CA 91109, USA\\
$^{39}$ Department of Astronomy, University of Michigan, Ann Arbor, MI 48109, USA\\
$^{40}$ Department of Physics, University of Michigan, Ann Arbor, MI 48109, USA\\
$^{41}$ Department of Physics, Cornell University, Ithaca, NY 14853, USA\\
$^{42}$ Instituto de Fisica Teorica UAM/CSIC, Universidad Autonoma de Madrid, 28049 Madrid, Spain\\
$^{43}$ Department of Astronomy and Steward Observatory, University of Arizona, Tucson, AZ 85721\\
$^{44}$ School of Physics, University of Melbourne, Parkville VIC 3010, Australia\\
$^{45}$ Department of Physics, ETH Zurich, Wolfgang-Pauli-Strasse 16, CH-8093 Zurich, Switzerland\\
$^{46}$ Institute for Advanced Study, Princeton, NJ 08540, USA\\
$^{47}$ Center for Computational Astrophysics, Flatiron Institute, New York, NY 10010, USA\\
$^{48}$ Santa Cruz Institute for Particle Physics, Santa Cruz, CA 95064, USA\\
$^{49}$ Center for Cosmology and Astro-Particle Physics, The Ohio State University, Columbus, OH 43210, USA\\
$^{50}$ Department of Physics, Florida State University, Keen Physics Building, Tallahassee, Florida 32306, USA\\
$^{51}$ Department of Physics and Astronomy, Rutgers University, Piscataway, NJ 08854-8019, USA\\
$^{52}$ Harvard-Smithsonian Center for Astrophysics, Cambridge, MA 02138, USA\\
$^{53}$ Lawrence Berkeley National Laboratory, 1 Cyclotron Road, Berkeley, CA 94720, USA\\
$^{54}$ Australian Astronomical Optics, Macquarie University, North Ryde, NSW 2113, Australia\\
$^{55}$ Departamento de F\'isica Matem\'atica, Instituto de F\'isica, Universidade de S\~ao Paulo, CP 66318, S\~ao Paulo, SP, 05314-970, Brazil\\
$^{56}$ George P. and Cynthia Woods Mitchell Institute for Fundamental Physics and Astronomy, and Department of Physics and Astronomy, Texas A\&M University, College Station, TX 77843,  USA\\
$^{57}$ Instituto de Astrof\'isica and Centro de Astro-Ingenier\'ia, Facultad de F\'isica, Pontificia Universidad Cat\'olica de Chile, Av, 7820436 Macul, Santiago, Chile\\
$^{58}$ Instituci\'o Catalana de Recerca i Estudis Avan\c{c}ats, E-08010 Barcelona, Spain\\
$^{59}$ Physics Department, University of Milano-Bicocca, Piazza della Scienza, 3 - 20126 Milan, Italy\\
$^{60}$ Department of Physics, Yale University, New Haven, CT 06520, USA\\
$^{61}$ Department of Physics and Astronomy, Haverford College, Haverford, PA 19041, USA\\
$^{62}$ Center for Astrophysics and Space Astronomy, Department of Astrophysical and Planetary Science, University of Colorado, Boulder, CO 80309, USA\\
$^{63}$ NASA Postdoctoral Program Senior Fellow, NASA Ames Research Center, Moffett Field, CA 94035, USA\\
$^{64}$ Department of Physics and Astronomy, Pevensey Building, University of Sussex, Brighton, BN1 9QH, UK\\
$^{65}$ School of Physics and Astronomy, University of Southampton,  Southampton, SO17 1BJ, UK\\
$^{66}$ Cerro Tololo Inter-American Observatory, National Optical Astronomy Observatory, Casilla 603, La Serena, Chile\\
$^{67}$ Brandeis University, Physics Department, 415 South Street, Waltham MA 02453\\
$^{68}$ Instituto de F\'isica Gleb Wataghin, Universidade Estadual de Campinas, 13083-859, Campinas, SP, Brazil\\
$^{69}$ Department of Astronomy and Astrophysics, University of Toronto, 50 St. George St., Toronto, ON M5S 3H4, Canada\\
$^{70}$ Computer Science and Mathematics Division, Oak Ridge National Laboratory, Oak Ridge, TN 37831\\
$^{71}$ Institute of Cosmology and Gravitation, University of Portsmouth, Portsmouth, PO1 3FX, UK\\
$^{72}$ NASA/Goddard Space Flight Center Observational Cosmology Laboratory, Greenbelt, MD 20771, USA\\

\appendix

\section{Prior distribution of the halo model}
We hereby present the distributions of the splashback radius, $r_{\rm sp}$, and the corresponding logarithmic slope of the density profile at $r_{\rm sp}$, drawn from the prior distribution of our halo model (Table~\ref{tab:modeling_parameters}). 

Among the halo density profiles drawn from the prior distribution, we remove the ones for which the logarithmic derivative is monotonically increasing/decreasing in $r$ (no splashback-like feature). It amounts to $\sim$20\% of the full prior distribution, which indicates our model is flexible enough to generate profiles without any splashback-like feature. 

\begin{figure}
\centering
\includegraphics[width=1.0\linewidth]{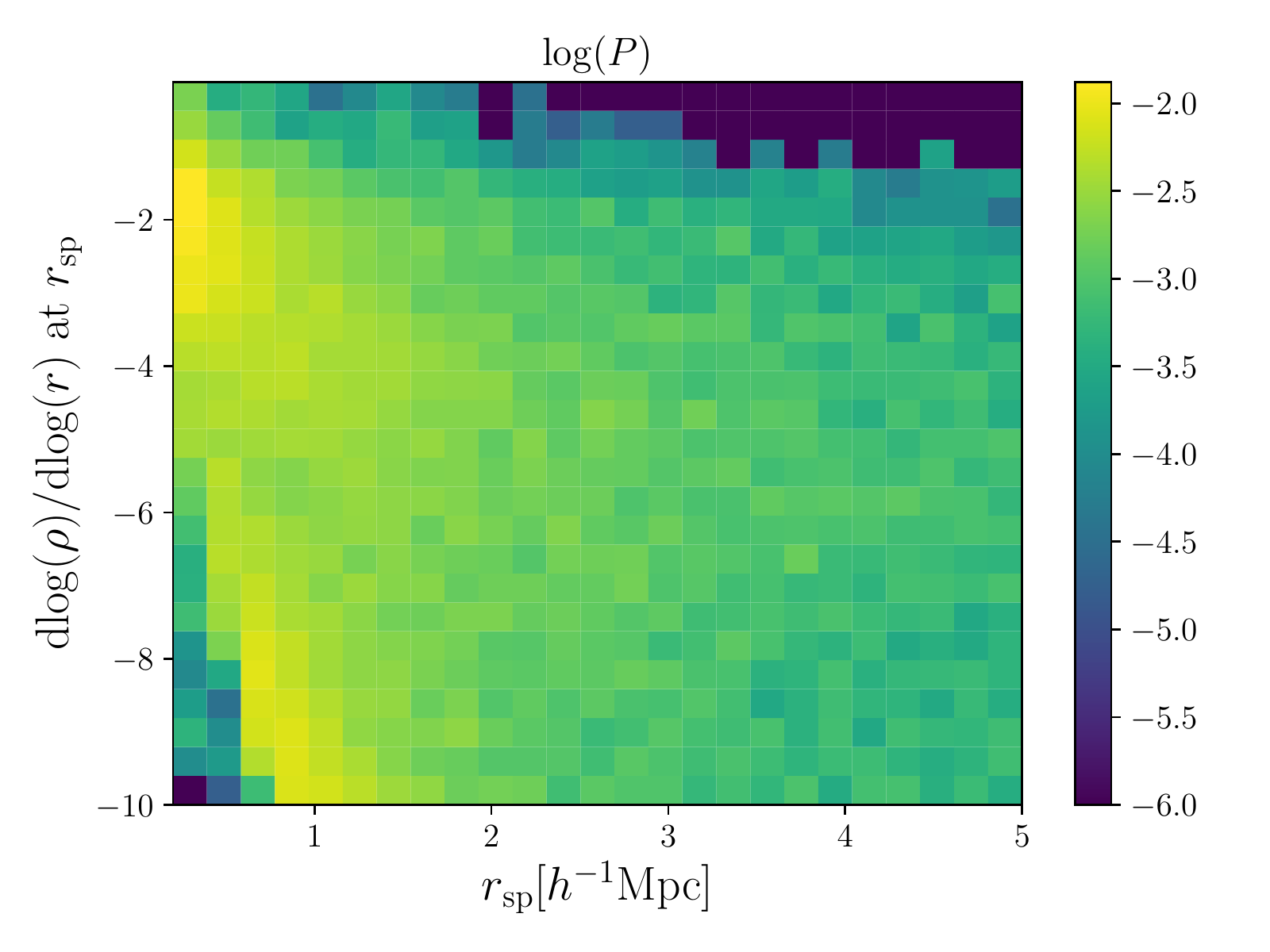}
\caption{The 2-dimensional probability distribution of $r_{\rm sp}$ (x-axis) and the logarithmic derivative of the density profile at $r_{\rm sp}$ (y-axis), drawn from the prior distribution of the halo model (Table~\ref{tab:modeling_parameters}).
}
\label{fig:prior_model}
\end{figure}

The result is shown in \Fref{fig:prior_model} in terms of the probability density. The x-axis represents the location of $r_{\rm sp}$ and the y-axis the corresponding logarithmic slope at $r_{\rm sp}$. One can see that the prior distribution covers the entire space fairly evenly.

\end{document}